\documentclass[12pt]{article}

\usepackage{verbatim}
\usepackage{lscape}
\usepackage{multirow}
\usepackage{amssymb,amsmath,amsthm}
\usepackage{amstext}
\usepackage{amscd}
\usepackage{graphics}
\usepackage{epsfig}
\usepackage[mathscr]{eucal}
\usepackage{times}
\usepackage{srcltx}
\usepackage[normalem]{ulem}
\usepackage{bbold}
\usepackage{shadow}
\usepackage[backref]{hyperref}

\def\wt{\widetilde}
\newcommand{\Rho}{{\mbox{\sf P}}}

\def\sss{\scriptscriptstyle}
\def\Real{\mathbb R}

\def\sss{\scriptscriptstyle}
\def\d{\partial}

\def\be{\begin{equation}}
\def\ee{\end{equation}}
\def\beq{\begin{equation}}
\def\eeq{\end{equation}}
\def\bea{\begin{eqnarray}}
\def\eea{\end{eqnarray}} 

\def\eqn#1{(\ref{#1})}

\def\nn{\nonumber}

\def\sideremark#1{\ifvmode\leavevmode\fi\vadjust{\vbox to0pt{\vss
 \hbox to 0pt{\hskip\hsize\hskip1em
 \vbox{\hsize3cm\tiny\raggedright\pretolerance10000
  \noindent #1\hfill}\hss}\vbox to8pt{\vfil}\vss}}}

\renewcommand{\tilde}{\widetilde}
\renewcommand{\hat}{\widehat}

\newcommand{\bref}[1]{\textbf{\ref{#1}}}

\newcommand{\gh}[1]{\mathrm{gh}(#1)}

\newcommand{\dd}{\partial}
\renewcommand{\d}{\partial}

\renewcommand{\geq}{\,{\geqslant}\,}
\renewcommand{\leq}{\,{\leqslant}\,}

\newcommand{\binner}[2]{%
  {\langle}\kern-4.15pt{\langle}#1{,}\,#2{\rangle}\kern-4.15pt{\rangle}}
\newcommand{\commut}[2]{[#1{,}\,#2]}

\newcommand{\half}{\mathchoice{%
    \ffrac{1}{2}}{\frac{1}{2}}{\frac{1}{2}}{\frac{1}{2}}}

\newcommand{\ffrac}[2]{\raisebox{.5pt}%
  {\footnotesize$\displaystyle\frac{#1}{#2}$}\kern1pt}

\newcommand{\red}{\mathrm{red}}

\newcommand{\dl}[1]{\frac{\dd}{\dd #1}}

\def\const{\mathop\mathrm{constant}\nolimits}

\newcommand{\bundle}[1]{\mathbf{#1}}
\newcommand{\manifold}[1]{\mathscr{#1}}

\newcommand{\manM}{\manifold{M}}

\newcommand\IcdotD{{I\!\cdot\!D}}

\newcommand{\Liealg}{\mathfrak}

\newcommand{\algp}{\Liealg{p}}

\def\cA{\mathcal{A}}

\def\cE{\mathcal{E}}
\def\cF{\mathcal{F}}
\def\cG{\mathcal{G}}
\def\cH{\mathcal{H}}

\def\cO{\mathcal{O}}

\def\grad{{\rm Grad}}
\def\div{{\rm Div}}
\def\tr{{\rm Tr}}
\def\cA{{\mathcal A}}

\def\Grad{\mathfrak{Grad}}
\def\ID{{\IcdotD}}
\def\Div{\mathfrak{Div}}
\def\IDZ{\mathfrak{\wt\Div}}
\def\Tr{\mathfrak{Tr}}

\def\ngh{N_{\rm ghost}}

\def\BGST{Barnich:2004cr}
\def\BGadS{Barnich:2006pc}

\def\AG{Alkalaev:2009vm}

\def\GoverWal{Gover:2008sw}
\def\GW{Gover:2009vc}

\numberwithin{equation}{section} \makeatletter
\begin{document}
{~~~}
\vspace{-1cm}

\begin{flushright}\small
FIAN-TD-2011-05
\end{flushright}

\thispagestyle{empty}

\vspace{2.2cm}
\setcounter{footnote}{0}
\begin{center}
\vspace{-25mm}
{\Large
 {\bf Massive Higher Spins from BRST and Tractors}\\[5mm]

 {\sc \small
     Maxim Grigoriev$^{\mathfrak G}$  and Andrew Waldron$^{\mathfrak W}$\\[4mm]
            {\em\small~${}^{\mathfrak G}\!$
            Tamm Theory Department,\\ Lebedev Physics Institute, Leninsky prospect 53, 119991 Moscow, Russia\\
            {\tt grig@lpi.ru}\\[4mm]  
                       ~${}^{\mathfrak W}\!$
            Department of Mathematics\\ 
            University of California,
            Davis CA 95616, USA\\[-2mm]
            {\tt wally@math.ucdavis.edu}
            }}

 }

\bigskip

{\sc Abstract}\\[-4mm]
\end{center}

{\small
\begin{quote}

We obtain the higher spin tractor equations of motion conjectured by Gover {\it et al.} from a 
BRST approach and use those methods to prove that they describe massive, partially massless
and massless higher spins in conformally flat backgrounds. The tractor description makes invariance
under local choices of unit system manifest. In this approach, physical systems are described by conformal, 
rather than (pseudo-)Riemannian geometry. In particular masses become geometric quantities, namely the weights
of tractor fields. Massive systems can therefore be handled in a unified and simple manner mimicking the gauge principle usually employed for massless models. 
{}From a holographic viewpoint, these models describe both the bulk and boundary theories in terms
of conformal geometry. This is an important advance, because tying the boundary conformal structure to that of the bulk theory gives greater control over a bulk--boundary correspondence.

\bigskip

\bigskip

\end{quote}
}

\newpage

\tableofcontents

\section{Introduction}

A fundamental principle is that physical descriptions of spacetime theories should not depend on any
local choice of coordinate system. This simple idea led Einstein to formulate the theory of gravity
in terms of (pseudo-)Riemannian geometry. If in addition, one adopts an old principle dating back to Weyl, that
physics also should be independent of local choices of unit systems, the gravitational coupling (the Planck mass)
must be elevated to a field which we shall call the ``scale'' (whose relative values at differing spacetime points expresses relative changes of unit systems, {\it i.e.} a non-dynamical dilaton or Weyl compensator) 
and Riemannian geometry is replaced by conformal geometry: Namely the theory of conformal equivalence classes
of metrics
\begin{equation}
[g_{\mu\nu}]=[\Omega(x)^2\ \! g_{\mu\nu}]\, ,
\end{equation} identified by equivalence up to Weyl transformations. In a recent series of works~\cite{Gover:2008pt,Gover:2008sw,Shaukat:2009hp,Shaukat:2010vb}, it has been pointed out that
a certain mathematical calculus coming from conformal geometry called ``tractor calculus''~\cite{BEG}
(which can be traced back to ideas of Thomas early last century~\cite{Thomas}, for further details see~\cite{Eastwood,Gover,Gover1,Gover:2002ay,Cap:2002aj})
can be employed to make manifest the local invariance of physical theories under choices of unit systems. 

In standard descriptions of higher spin theories, spin is encoded by the tensor structure of
the field content of the model (so is in some sense kinematic) while masses correspond to eigenvalues
of the second Casimir of an underlying isometry group (so, in turn, some form of Laplace operator). 
Of course, spin can be expressed in terms of quartic Casimir operators, but it would actually be more desirable
if both masses and spin could be formulated in terms of kinematics alone, independent of any precise details
of background isometries. This is precisely what happens in the tractor approach where $d$-dimensional Lorentz tensors
$\varphi_{\mu_1\ldots\mu_s}(x)$ (where $\mu=0,\ldots d-1$)
are replaced by $(d+2)$-dimensional tractor tensors $\varphi_{M_1\ldots M_s}$ (with $M=0,\ldots, d+1$) defined as sections of certain weighted tractor tensor bundles over $d$-dimensional spacetime. The weights of these bundles correspond to masses of physical fields\footnote{In four dimensions then, one still employs fields depending on four space time coordinates, but which are grouped into
multiplets under Weyl transformations labeled by six-dimensional indices. A simple example of this phenomenon is the four-velocity $v^\mu=\dot x^\mu$ which, along with the four-acceleration $a^\mu=\frac{\nabla v^\mu}{dt}$, forms a weight~1 tractor vector $V_M=\big( 0,\  v_m, \  \frac{v.a}{v.v})$.}.

For the case of massive higher spin theories, it was shown in~\cite{Gover:2008pt,\GoverWal}, that spin~$s$,
totally symmetric, massive, massless and partially massless theories could be formulated in terms of weight~$w$, rank~$s$, symmetric tractor tensors $\Phi_{M_1\ldots M_s}$. Moreover,  
Fronsdal--Curtright``-esque''
 gauge invariant equations of motion~\cite{Fronsdal:1978rb,Curtright:1979uz} were conjectured for these (generically massive) fields.
 Masses in these theories were expressed in terms of their weights $w$,
which could be viewed as the response of the system to changes of scale. Special, integer choices of weights $w=-1,0,\ldots,s-2$ 
and $w=s-2$ then corresponded, respectively,  to partially massless and strictly massless theories. This conjecture was checked for $s\leq 2$. Its key ingredient was an extremely simple formulation of the equations of motion in terms of: (i) Tractor generalizations of the de Wit--Freedman higher spin curvatures,
\begin{equation}
\Gamma^R_{M_1\ldots M_s} = s D_{(M_1} \Phi^R{}_{M_2\ldots M_s)} - D^R \Phi_{M_1\ldots M_s}\, ,
\end{equation} 
where $D^R$, stands for the Thomas $D$-operator, a Weyl covariant multiplet of operators generalizing the standard gradient operator (see Section~\bref{tractors} for details). (ii) A uniform coupling to scale requiring the higher spin curvatures to be orthogonal to the ``scale tractor'' $I^M$: 
\begin{equation}
I_R \, \Gamma^R_{M_1\ldots M_s} = 0\, .
\end{equation} 
This equation deserves a detailed explanation. The full set of tractor equations required to describe massive higher
spins are given in Figure~\ref{tequations}, but this one encodes the dynamics of the system and is a rewriting of the first equation in the Figure, while the others are constraints. The tractor higher spin curvatures are Weyl covariant so all breaking of Weyl invariance is through the scale tractor $I^M$. The scale tractor  is an extremely interesting quantity. It contains all information of the varying Planck mass, its length $I^2$ encodes the value of the cosmological constant, extremizing its square (as the integrand of an action principle) yields Einstein's equations and it is (tractor) parallel precisely when 
the class of metrics~$[g_{\mu\nu}]$ is conformally Einstein.

The main aim of this paper is to prove the conjectured  tractor description of massive higher spins of~\cite{Gover:2008pt,Gover:2008sw}. In fact we give both a direct demonstration that the equations of Figure~\ref{tequations} describe massive higher
spins in constant curvature spaces, as well a second proof relying on BRST methods. Although the BRST machinery requires the
introduction of extensive technology, the payback is a parent formulation which overlies the various approaches to massive higher spins.
Therefore this paper connects various approaches to massive higher spins. It is also intimately related to recent approaches to conformally invariant higher spin systems using a $(d+2)$-dimensional fiber over a $d$-dimensional base manifold~\cite{Bekaert:2009fg}. 

The equations of motion for a massive (totally symmetric)  spin~$s$ field in what we shall term an ``on-shell'' form
$$
(\nabla^2 - \mu^2) \phi_{\mu_1\ldots \mu_s} = 0 = \nabla^{\mu}\phi_{\mu\mu_2\ldots\mu_s}=\phi^\mu{}_{\mu\mu_3\ldots\mu_s}\, ,
$$
appear simple enough, but this simplicity hides a myriad of subtleties. Firstly one must decide how the eigenvalue~$\mu$ of the Laplacian is related to mass, and moreover exactly what mass means when one moves beyond the umbrella of the standard intuition
afforded by Minkowski isometries~\cite{Deser:1983mm,Deser:1983tm,Higuchi:1986py,Brink:2000ag}. Indeed, the above equations enjoy distinct, ``residual'' gauge invariances at $s$ different
values of the parameter~$\mu^2$. These correspond to so-called partially massless theories~\cite{Deser:2001us,Deser:2001pe,Deser:2001wx,Deser:2001xr,Deser:2003gw,Deser:2004ji}.
To write these equations in a form that follows from an action principle requires the
addition of one or another system of auxiliary fields.

There are various descriptions of massive higher spins in constant curvature spaces, all with their own peculiar advantages:
If one is interested in writing actions for massive higher spins in constant curvature spaces, cases of low lying values of~$s$ can easily be handled by generalizing the Hagen and Singh result~\cite{Singh:1974qz} for Minkowski space (see for example~\cite{Chang:1967zz,
Berends:1979rv,Polishchuk:1999nh,Deser:2001us,Deser:2001pe}). If manifest  covariance is jettisoned, the actions in a $3+1$ ADM-type decomposition~\cite{Arnowitt:1962hi} for physical helicities for all spins follow trivially from the analysis of~\cite{Deser:2001wx}. 

As far as general massive higher spin fields are concerned a representation-theoretic analysis and a related light-cone
description (which is very useful for holographic considerations) was given in~\cite{Metsaev:1999ui} (see also~\cite{Metsaev:2003cu,Metsaev:2004ee}). When an all spin, covariant approach is desired, the introduction 
of extra St\"uckelberg auxiliaries is highly expeditious, see~\cite{Zinoviev:2001dt}. The origin of these extra auxiliary fields can be easily
understood by reducing the massless theory in a flat space of one higher dimension along a conformal isometry~\cite{Biswas:2002nk}.
This method leads to generating functions for massive constant curvature higher spin actions in a minimal field content~\cite{Hallowell:2005np}. It is also possible to reduce from massless theories in constant curvature space to massive theories in a lower dimensional constant curvature space~\cite{Metsaev:2000qb,Artsukevich:2008vy}. Another approach that emphasizes the higher spin geometry of massive theories is to consider frame-like 
formulations~\cite{Zinoviev:2008ve,Zinoviev:2009gh,Ponomarev2010}. The constraint structure of massive higher spins 
can be better understood from a BRST perspective~\cite{Buchbinder:2006ge,Buchbinder:2007ix} (see also \cite{Buchbinder:2008ss,Bekaert:2003uc,Fotopoulos2009}). Another approach, quite closely related to the tractor description, relies on an AdS/CFT strategy to write actions, see~\cite{Metsaev:2009hp}.

In this Article  we focus on
understanding the {\it r\^ole} played by bulk conformal geometry for the description of massive higher spins.
Recently Metsaev has begun an investigation of the bulk-boundary aspects of massive constant curvature higher spins
based on conformal currents arranged in what amount to tractor multiplets~\cite{Metsaev:2009ym,Metsaev:2010zu}.
The results here, relying on the bulk conformal structure, coupled with those recent result for boundary conformal currents
will allow a description of massive excitations in which the bulk conformal structure determines the boundary conformal structure.
Very recently, it has been shown that this method allows solutions to be computed from a simple solution generating algebra~\cite{Gover:2011rz}.

The Paper is structured as follows. In Section~\bref{tractors} we review basic concepts from conformal geometry and explain
the main ingredients of tractor calculus needed for this work. We also connect these ideas to compensator methods employed
in the physics literature to describe conformal systems. In Section~\bref{radial} we review how massive higher spins in 
a $d$-dimensional 
constant curvature space can be obtained from their massless counterparts in $(d+1)$-dimensional flat space. There we also explain
how to represent these fields living on slices of the light cone in a $(d+2)$-dimensional ambient space. The relationships between
these representations of massive higher spins yield a brute force demonstration that the equations in Figure~\ref{tequations}
describe massive higher spins. In Section~\bref{BRST} we uncover the origin of the tractor higher spin equations using first quantized BRST machinery. This leads to a second proof of the conjecture. We then extend this to a parent BRST formulation which: gives a third proof of this; neatly explains the appearance of the Thomas-$D$ operator in the gauge transformations of the theory; underlies
an unfolded treatment of massive higher spins; and acts as a parent description from which the various BRST descriptions
follow.

 \begin{figure}
 \begin{center}
 \shabox{
 \begin{tabular}{c}
 Equations of Motion\\[3mm]
 $\Big(\IcdotD - \frak{Grad}\, \wt{\frak{Div}}\Big)\Phi=0$\\ $\Updownarrow$ \\[1mm]
 $\IcdotD\, \Phi_{M_1\ldots M_s}-sD_{(M_1}I^M\Phi_{M_2\ldots M_s)M}=0$\\[8mm]
 $\Big({\frak{Div}}-\frac12\, \frak{Grad}\, \frak{Tr}\Big)\Phi=0$\\ $\Updownarrow$ \\[1mm]
 $s D^M\Phi_{MM_2\ldots M_s}-\frac{s(s-1)}2D_{(M_2}\Phi^M{}_{M_3\ldots M_s)M}=0$\\[8mm]
 $\frak{Tr}^2\Phi=\wt{ \frak{Div}}\, \frak{Tr} \ \Phi=\wt{\frak{Div}}{\ \!}^2\,  \Phi=0$\\$\Updownarrow$ \\[1mm]
 $\qquad s(s-1)(s-2)(s-3)\Phi^M{}_M{}^N{}_{NM_5\ldots M_s}=0\qquad $\\[3mm]
 $s(s-1)(s-2)
 I^N\Phi^M{}_{MNM_4\ldots M_s}=0$\\[3mm]
 $s(s-1)I^M I^N \Phi_{MNM_3\ldots M_s}=0$
 \\[7mm]
 Gauge Invariances\\[3mm]
 $\delta \Phi = \frak{Grad} \, \varepsilon\, ,\quad \frak{Tr}\, \varepsilon = 0 = \wt{\frak{Div}}\, \varepsilon$ \\[1mm]
 $\Updownarrow$ \\[1mm]
 $\delta \Phi_{M_1\ldots M_s}=D_{(M_1} \varepsilon_{M_2\ldots M_s)}$\\[2mm]
 $s(s-1)\, \varepsilon^M{}_{MM_3\ldots M_{s-1}}=0=s\, I^M\varepsilon_{MM_2\ldots M_{s-1}}$
 \end{tabular}
 }
 \end{center}
 \caption{Tractor equations of motion unifying massive, massless and partially massless higher spins in standard and index-free notations.
 \label{tequations}}
 \end{figure}

\section{Conformal Geometry and Tractor Calculus}
\label{tractors}

Tractor calculus allows conformal geometry to be described in a way that keeps Weyl invariance,
\be
g_{\mu\nu}\mapsto \Omega^2 g_{\mu\nu}\, ,\label{gW}
\ee
manifest at all times. Moreover, by coupling to a choice of scale ({\it i.e.}, a dilaton field/spacetime-dependent Planck mass encoding how local choices of unit systems
vary from spacetime point to point), it provides a natural framework for describing any physical system, in particular those with dimensionful  couplings that are apparently {\it not} Weyl invariant~\cite{Gover:2008pt,\GoverWal,Shaukat:2009hp,Shaukat:2010vb}, in a way manifestly independent of local choices of unit systems. Although, tractors naturally describe {\it arbitrary} conformal classes of metrics $[g_{\mu\nu}]=[\Omega^2 g_{\mu\nu}]$ on a manifold $\manM$, they are
perhaps  most easily understood by examining the canonical flat model for conformal geometries, namely the conformal sphere.

\subsection{The Flat Model for Conformal Geometry}
\label{flat bundle}
\begin{figure}
 \begin{center}
 \includegraphics[height=5cm,width=6cm]{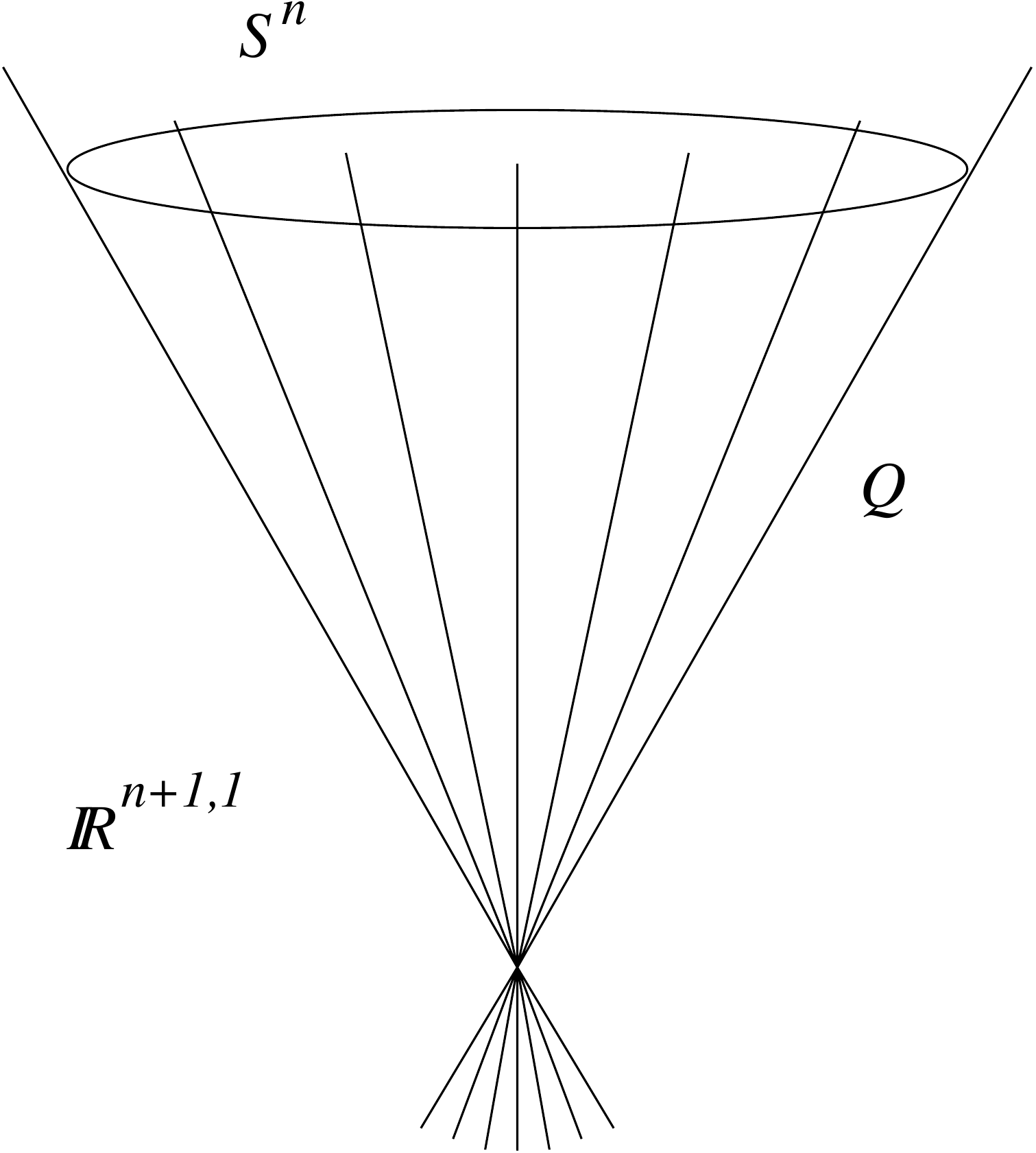}
 \end{center}
 \caption{\label{rays} The flat conformal space~$\manM$ in Riemannian signature realized as null rays.}
 \end{figure}

The conformal sphere is the conformal manifold $(\manM,[g_{\mu\nu}])=(\Real^n,[\delta_{\mu\nu}])$ whose  class of metrics
has Riemannian signature. This space is modeled on rays in the lightcone of a $(n+1,1)$-dimensional, ambient,  Minkowski space; see Figure~\ref{rays}. 
For physical applications, Lorentzian signatures are more relevant. In this setting 
the flat model of conformally Minkowski space~$(\Real^d,[\eta_{\mu\nu}])$ is  described in terms of a~$(d,2)$-dimensional ambient pseudo-Riemannian space\footnote{Most formul\ae\  in this paper may be transcribed to an arbitrary metric signature.}
with metric
\bea
 ds^2&=&  \eta_{mn}dY^m dY^n-(dY^d)^2+(dY^{d+1})^2=dY^A dY_A+(dY^{d+1})^2\nn\\[3mm]
         &=&\ 2\, dY^+ dY^- + \eta_{mn}dY^m dY^n \ = \ \eta_{MN} dY^M dY^N\, .
 \label{flatamb}
\eea
Here~$Y^M=(Y^m,Y^\pm)$, $Y^A=(Y^m, Y^d)$ and $\eta_{mn}$ is the standard $(d-1,1)$ Minkowski metric.
The conformal space~$\manM$ is the space of rays in the ambient space null cone~$Y^2=0$. The ambient space orthogonal group~$G=O(d,2)$ obviously acts
transitively on the  space~$Q$ of null rays and this gives the coset representation~$\manM=O(d,2)/P$ of the conformal space~$\manM$, where~$P$ is the parabolic subgroup preserving a given null ray.

If we view the equivalence relation on the cone 
$$
Y^M\sim\Omega Y^M\, ,\quad \Omega\in {\mathbb R}^+\, ,
$$
as a gauge equivalence, then by a fixing a gauge the conformal space~$\manM$ can then be identified as a submanifold of the cone. This submanifold is canonically  equipped with a metric in the conformally flat equivalence class of metrics induced from the ambient metric. Varying this gauge choice  therefore generates the conformal class of metrics belonging to~$\manM$.  Standard gauge choices are:
$Y^{d+1}=1$ which (together with~$Y^2=0$) gives the defining equation for anti de Sitter (AdS) space;~$Y^+=1$ giving the flat space representation
of the conformal space; and $Y^d=1$ giving de Sitter (dS) space. All these metrics are, of course, conformally equivalent.

The conformal space is equipped with a flat Cartan connection which can be obtained by pulling back the canonical Maurer--Cartan form
on~$G$ to the coset (seen as a submanifold of~$G$). From the viewpoint of tractor calculus, this connection is precisely the (flat) tractor connection.

\subsection{General Tractor Bundle}
\label{gen-trac}

The general  weight $w$,  tractor (vector)  bundle ${\mathcal E}^M[w]$, is a rank $d+2$ bundle over 
a $d$-dimensional manifold $\manM$ built from
$${\mathcal E}^M[w]={\mathcal E}[w+1]\oplus{\mathcal E}^m[w]\oplus{\mathcal E}[w-1]\, .$$
Here ${\mathcal E}[w]$ denotes the weight $w$  bundle of conformal densities over $\manM$
and the  abstract index $m$ is used to denote (soldered) vector-valued conformal densities.
The tractor bundle is defined by the transformation of sections $T^M\in\Gamma{\mathcal E}^M[w]$
under conformal rescalings
\begin{equation}
 \begin{pmatrix}T^+\\[1mm]T^m\\[1mm]\ T^-\ \end{pmatrix}\mapsto \Omega^w
\begin{pmatrix}\Omega T^+\\[1mm]T^m+\Upsilon^m T^+\\[1mm]\ \Omega^{-1}\big(T^--\Upsilon_mT^m-\half \Upsilon_m\Upsilon^m\,  T^+\big)\ \end{pmatrix}
\, ,\label{tractor gauge}
\end{equation} 
where $\Omega\in C^\infty(\manM)$ and\footnote{In fact this definition is often stated without reference to a metric 
by replacing soldered vectors by tangent vectors so that ${\mathcal E}^M[w]={\mathcal E}[w+1]\oplus T\manM[w-1]\oplus{\mathcal E}[w-1]$ and $\Upsilon =\Omega^{-1}\, d\Omega\in \Gamma T^*M$.} 
\begin{equation}
\Upsilon^m:=e^{\mu m}\Omega^{-1}  \d_\mu \Omega\, 
\end{equation}   
(where $e^{\mu m}$ is the inverse vielbein).
Although this definition, due to~\cite{BEG} may at first appear arcane, it is actually the canonical way to handle conformal geometries and underlies a comprehensive conformal calculus.
Moreover, the bundle ${\mathcal E}^M[0]$ admits a natural connection underlying an intimate relationship
between conformal and Einstein structures. In plain language, a manifold admits a metric
conformal to an Einstein metric whenever there is a parallel tractor. To better understand
this distinguished  connection and its relation to other approaches, let us briefly backtrack to 
bundles whose structure group is the full conformal group.

Consider a principal $G=O(d,2)$ bundle over~$\manM$: 
Let $\bundle{V}(\manM)$ be an associated pseudo-Riemannian vector bundle with structure group~$G$ and $\eta$ the fiberwise inner product. The conformal structure is defined in terms of the~${\mathfrak o}(d,2)$ connection $\nabla=d+{\mathcal A}$ and a null line subbundle~$\bundle{X}$ generated by a section $X$ (defined modulo rescalings) subject to $X^2:=\eta(X,X)=0$.
The choice of~$\bundle{X}$ singles out a parabolic subgroup in~$P\subset G$ as a stability subgroup of the~$X$ ray. 
We now impose the following conditions on this data
\begin{enumerate}
\item \label{maxrank}$\nabla X$ understood as a fiberwise map from $T\manM$  to $\bundle{V}(\manM)$ has vanishing kernel
and $\eta$ restricted to its image is non-degenerate. 
\item \label{para}The curvature $F=d\cA+\cA\wedge\cA\ $ belongs to the Lie algebra $\algp$ of $P$.
\item\label{orthogonal} $F(T\manM^\perp)\subset T\manM$, where $T\manM^\perp$ denotes
an orthogonal complement of $T\manM$  (the image of $\nabla X$) in $\bundle{V}$.
\end{enumerate}

Using these conditions it is easy to check that there exists a local frame $(E_+$, $E_-$, $E_m)$
such that $E_-$ generates $\bundle{X}$, $\eta({E_+},{E_-})=1$ and $({E_+}$, ${E_-})$ are basis elements in $T\manM^\perp$, while $\eta({E_m},{E_n})=\eta_{mn}$  and $E_m$ form a  basis for $T\manM \in \bundle{V}$.
Moreover, condition~\ref{orthogonal} implies that the local frame can be chosen such that the corresponding connection components  satisfy $\cA^+{}_+=\omega^-{}_-=0$. To see this, observe that 
$\alpha=\eta(\nabla E_-,E_+)$ is a closed one form thanks to~\ref{orthogonal} so that
$\alpha=d\lambda$ and one can therefore redefine  basis sections $E_+,E_-$ according to $E_+ \to \exp(\lambda) E_+ $ and $E_- \to \exp(-\lambda) E_- $ to achieve $\cA^+{}_+=\cA^-{}_-=0$.
Writing out the non-vanishing connection components in matrix form, we have:
\be\label{covder}
\nabla_\mu = {\mathbb 1}\,  \partial_\mu +\begin{pmatrix}0&-e_{\mu n} & 0 \\[1mm] \Rho_\mu{}^m&\omega_\mu{}^m{}_n&e_\mu{}^m\\[1mm]0&-\Rho_{\mu n}&0\end{pmatrix}\, .
\ee
Here we have used condition~\ref{maxrank} to identify $\cA_\mu{}^m{}_-$ with the 
vielbein $e_\mu{}^m$.
 Calling $\cA_\mu{}^m{}_n=\omega_\mu{}^m{}_n$, it follows from condition~\ref{para} that $de^m+\omega^m{}_n \wedge e^n=0$ and hence~$\omega^m{}_n$ is the torsion-free Lorentz
connection. Finally, the  tensor $\Rho_\mu{}^m$ can be identified with the so-called Schouten or \textit{rho}-tensor\footnote{Recall that the Schouten tensor is the trace-adjusted Ricci tensor defined by $$R_{\mu\nu\rho\sigma}=W_{\mu\nu\rho\sigma}+\Rho_{\mu\rho}g_{\nu\sigma} 
-\Rho_{\nu\rho}g_{\mu\sigma}
-\Rho_{\mu\sigma}g_{\nu\rho}
+\Rho_{\nu\sigma}g_{\mu\rho}
\, ,$$ where $W_{\mu\nu\rho\sigma}$ is the trace-free Weyl tensor. We denote $\Rho:=\Rho_\mu^\mu=\frac{R}{2(d-1)}$.}
associated with the Riemannian metric $g_{\mu\nu}=\eta_{mn}e_\mu{}^m e_\nu{}^n$ (or, more
invariantly, $ds^2=\eta(\nabla E_-,\nabla E_-)$).  This identification can be justified as follows:
the $m,n$ component of the curvature  $F^m{}_n$ has the structure $R^m{}_n-e^m\wedge \Rho_n-\Rho^m\wedge e_n$, where the two-form $R^m{}_n$ is usual Riemann tensor. At the same time it vanishes in the conformally flat case and hence can be identified with the Weyl tensor.

The tractor bundle is precisely the one for which the  structure group is reduced to $P$ in order to preserve
the subbundle $\bundle{X}$. In addition the allowed gauge transformations are restricted to those preserving the structure of the connection. 
More precisely,
requiring the $\algp$-valued gauge parameter $\lambda^M{}_N$ to preserve the condition $\cA^+{}_+=0$
algebraically determines $\lambda^m{}_+$ in terms of $\lambda^+{}_+$. In this way the residual gauge transformations are parametrized by $\lambda^+{}_+$ and $\lambda^m{}_n$. Those with
$\lambda^+{}_+=0$ are identified with local Lorentz transformations while those with
$\lambda^m{}_n=0$ with local Weyl transformations. Writing these out in matrix terms 
gives precisely the defining formula~\eqn{tractor gauge} (at weight $w=0$).
More general tractor bundles can be obtained by considering tensor powers of the tractor bundle ${\cal E}^M[w]$ and tensoring with the associated rank one conformal density bundles carrying weight $w$ representations of $P$.
Sections of these bundles are referred to as tractors. As we shall see, they naturally encode the field content
of physical systems and their weights will in fact  correspond to masses.

Of course,  for many applications, the original definition of the tractor bundle in terms of a direct sum of
the tangent bundle and a pair of line bundles, rather than constraints on curvatures, is the simplest approach.

In particular, in this  setup the subbundle $\bundle X$ is generated by the so-called canonical tractor $X=E_-$ which is 
a distinguished section of ${\mathcal E}[1]$. In a matrix notation, it is defined by 
\be
X^M=\begin{pmatrix}0\\[1mm]0\\[1mm]\ 1\ \end{pmatrix}\, .
\ee
It is also worth noting that although the tractor bundle was first introduced by Thomas early in the twentieth
century~\cite{Thomas,ThomasBook} (and could also be partly credited to Cartan~\cite{Cartan}), it also appeared (implicitly) in a physical context in the original construction of conformal gravity as a gauge theory of a the spacetime conformal group in~\cite{Kaku:1977pa,Kaku:1978nz,Kaku:1977rk,Townsend:1979ki}.
The above discussion, essentially follows the conformal compensator method for conformal gravity introduced in~\cite{Preitschopf:1997gi,Preitschopf:1998ei}. In particular, after gauge fixing the canonical tractor $X$ can be related to the conformal compensator.

A fundamental ingredient of tractor calculus, and probably the most basic structure (aside from the covariant derivative) on the tractor bundle is the Thomas~$D$-operator~$D^M$:
\begin{equation}\label{ThomD}
D^M =
\begin{pmatrix}
w(d+2w-2) \\[2mm]
(d+2w-2)\nabla^m \\[2mm]
-(g^{\mu\nu}\nabla_{\mu}\nabla_{\nu} +w\, \Rho)
\end{pmatrix}\, ,  \\%[15mm]
 \end{equation}
which maps weight~$w$, rank~$k$  sections of the tensor tractor bundle to weight~$w-1$,  rank~$k+1$ sections. 
It is important to note that the Thomas~$D$-operator (being second order in derivatives) does not obey a Leibnitz rule.
It is, however, always null
\be\label{null}
D^M D_M=0\, .
\ee

We are now ready to explain the relationship between parallel, weight zero tractors and conformally Einstein
metrics. Firstly, as advocated in the Introduction, we replace the gravitational coupling constant with a spacetime varying scale field $\sigma$---the gauge field measuring how unit systems vary locally. In a choice of gauge in which it is constant, it corresponds to the Planck mass via
$\sigma=\kappa^{\frac{2}{d-2}}$. Indeed, when working with conformal geometries,
specifying a choice of the weight~1 scalar field~$\sigma$ amounts to making a canonical choice of metric, since 
the double conformal equivalence class~$[g_{\mu\nu},\sigma]$ always has a representative~$[\sigma^{-2} g_{\mu\nu}, 1]$. The first slot is the canonical metric. Tautologically, asking this metric to be Einstein (clearly a highly desirable choice on physical grounds), implies that the conformal class of metrics is conformally Einstein.

{}Now, from the scale~$\sigma$, 
we can  define the so-called scale tractor
\begin{equation}\label{scale tractor}
 I^M=\frac 1d D^M \sigma=\begin{pmatrix}\sigma \\[2mm] \nabla^m\sigma\\[1mm] -\frac1d\big(\nabla^2\sigma +\Rho\sigma\big)\end{pmatrix}\, .
\end{equation}
To determine when the scale tractor is parallel we use the tractor covariant derivative~\eqn{covder} to compute
\be
\nabla_\mu I^M=\begin{pmatrix}0\\[1mm]
\big(\nabla_\mu\nabla^m-\frac1d e_\mu{}^m\nabla^2\big)\sigma+\big(\Rho_\mu^m-\frac1d e_\mu{}^m \Rho\big)\sigma\\[2mm]
-\frac1d\nabla_\mu\big(\nabla^2\sigma +\Rho\sigma\big)-\Rho_\mu^m \nabla_m\sigma \end{pmatrix}\, .
\ee
Examining this formula for the choice of Weyl frame $\sigma=1$  we immediately see that a conformal manifold is conformally Einstein when the scale tractor is parallel~\cite{BEG}.
Moreover, it follows that Thomas-$D$ operator commutes with the scale tractor on conformally Einstein manifolds
\be\label{DI}
 [D_M,I_N]=0.
 \ee
In fact the converse statement to above holds too; conformally Einstein manifolds admit a parallel scale tractor~\cite{BEG}.

For future reference, we note that in the conformally flat case (a stronger condition than the conformally Einstein one) Thomas-$D$ operators commute
\be\label{DD}
[D^M,D^N]=0\, .
\ee
This relation holds precisely when the tractor connection is flat  (in fact it is not difficult to verify that its curvature
is built from the Weyl and Cotton tensors).
The relations~\eqn{DI},~\eqn{null} and~\eqn{DD} will play an important {\it r\^ole} in our tractor construction of massive higher spin systems.

\subsection{Symmetric Tractor Tensors}

\label{OSC}

An efficient way to handle symmetric tensors of arbitrary rank is to view them as polynomial functions
of coordinates $z^\mu$ on the fibers of the tangent bundle over the space-time manifold. These can be also seen as functions of commuting coordinate differentials $dx^\mu$. For example, the metric tensor ${\rm g}(x^\mu,z^\mu)=g_{\mu\nu} z^\mu z^\nu$
is a quadratic polynomial in $z^\mu$. From a first quantized point of view, symmetric tensors are wavefunctions and the operators $\big(z^\mu,\frac{\partial }{\partial z^\mu}\big)$ are oscillators corresponding to spinning degrees of freedom.
It is then advantageous to introduce geometric operators mapping symmetric tensors to symmetric tensors such as
\begin{equation}
{\rm g} = g_{\mu\nu} z^\mu z^\nu\, ,\qquad {\rm N}=z^\mu\frac{\partial }{\partial z^\mu}\, ,\qquad {\rm tr}=g^{\mu\nu}
\frac{\partial }{\partial z^\mu}\frac{\partial }{\partial z^\nu}\, ,
\end{equation} 
\begin{equation}
{\rm grad}=z^\mu \nabla_\mu\, ,\qquad {\rm div}=\frac{\partial }{\partial z^\mu}\nabla^\mu\, .\label{ops}
\end{equation} 
The three operators on the first line respectively multiply symmetric tensors by the metric and symmetrize, 
count symmetric tensor indices and 
trace symmetric tensors. They obey an $\frak{sl}(2)$ algebra. The two operators on the second line perform the standard
geometric operations indicated by their names; they form a doublet under the adjoint action of the $\frak{sl}(2)$ generators.
On locally symmetric spaces, their commutator produces the Lichnerowicz wave operator, which is central under the above algebra
~\cite{Hallowell:2007zb} (in flat space, altogether, these then generate the $\frak{sl}(2)$ Jacobi algebra; when conformal isometries are present,
the generators corresponding to the negative roots in $\frak{sp}(4)$---of which the Jacobi algebra is a parabolic subalgebra---are also realized~\cite{Burkart:2008bq}).

It is particularly useful to define analogous operators on symmetric, tractor tensors
\be\label{calV}
\Phi=\sum_{s} \Phi_{M_1\ldots M_s} Z^{M_1}\cdots Z^{M_s}\, ,
\ee
where~$\Phi_{M_1\ldots M_s}$ is a weight~$w$, rank~$s$ totally symmetric, tractor tensor and the 
formal (commuting) variables~$Z^M$  are introduced as a bookkeeping device for tractor indices carry weight zero.

{}From the tractor metric $\eta^{MN}$, we can build an $\frak{sl}(2)$ triplet of operators analogous to the trace, index and metric  operators
above

\begin{equation}
\frak{G}=Z\cdot Z\, ,\qquad \frak{N}=Z\cdot \frac{\partial}{\partial Z}\, ,\qquad \frak{Tr}=\frac{\partial}{\partial Z}\cdot\frac{\partial}{\partial Z}\, .
\end{equation} 
Here and in what follows we use notation $A\cdot B$ to denote the invariant contraction of ambient space indexes.
There are now a pair of doublets built from the Thomas $D$-operator and scale tractor
\begin{equation}
\begin{aligned}
\Grad &= Z \cdot  D\, ,&\qquad \Div  &= \frac{\partial}{\partial Z} \cdot D\, ,\   \\
%\Grad =\ \  Z^M D_M\, ,\qquad \Div \ = \ \frac{\partial}{\partial Z^M} D^M\, ,\   \nn                                                                                                                                                        
\wt{\frak{\Grad}}&= Z\cdot  I\, ,&\qquad \IDZ &= \frac{\partial}{\partial Z} \cdot I \, .
\end{aligned} 
\end{equation} On conformally flat spaces $\Div$ and $\Grad$ commute so that when the scale tractor~$I^M$ is
tractor parallel ({\it i.e.} on conformally Einstein spaces) the only non-vanishing commutators are
\begin{equation}
[\Div,\wt{\frak{Grad}}]=\IcdotD = [\IDZ,\Grad]\, ,\qquad [\IDZ,\wt{\frak{Grad}}]=I^2\, ,
\end{equation} in which case both $\IcdotD$ and $I^2$ are central.

Of particular interest in this Article will be the first class constraint (Lie) algebra in Figure~\ref{tractor-const}.
\begin{figure}
\begin{center}
\shabox{
%\begin{gathered}
\begin{tabular}{c}
Constraints\\[2mm]
$
{\frak g}=
\big\{
\Grad,\  
\ID , \  
\frak{Div}, \ 
\IDZ ,\  
\Tr \ 
\big\}
$\\[6mm]
Algebra\\[2mm]
$[\IDZ,\Grad]=\IcdotD\, ,\qquad [\Tr,\Grad]=2\, \Div\, $
\end{tabular}
}
\end{center}
\caption{Tractor constraint algebra  on conformally flat spaces. The Lie algebra cohomology $H^1(\frak{g},{\cal V})$ describes
massive higher spins.
\label{tractor-const}}
\end{figure}
In fact our main aim is to show that the cocycle and coboundary conditions of its  degree one Lie algebra cohomology (a special case of Hamiltonian BRST quantization) amount to the equations of motion and gauge invariances, respectively,
of  the theory of massive higher spins on any conformally flat manifold.
Before discussing massive higher spins and BRST quantization, we show how to formulate 
this algebra on a $(d+2)$-dimensional flat ambient space.

\subsection{Ambient Space}\label{ambient space}

For many computations, the  curved ambient construction of tractor operators~\cite{GP,CG} is very useful. In particular, in~\cite{\GW}
it was shown that  for conformally flat spaces, the ambient space corresponded to the momentum space of $(d+2)$-dimensional 
massless scalar field. In this picture the null cone is the moduli space of massless excitations and the tractor operators
are the momentum space generators of the conformal symmetries of the model. 
This framework is also intimately related~\cite{Bonezzi:2010jr} to  Bars' two times approach~\cite{Bars:1997bz,Bars:2000qm}.
Let us briefly sketch the construction of the curved Fefferman--Graham construction and then specialize to the conformally flat case we mostly require in this Article.

A conformal structure on a manifold determines a Fefferman--Graham ambient metric which admits a hypersurface orthogonal
homothety. In the conformally flat case this homothety is generated by the Euler vector field. In the curved ambient construction,
the corresponding homothetic vector field (whose components $X^M$ play a {\it r\^ole} similar to coordinates, but from the perspective of Section~\bref{gen-trac} amount to the canonical tractor)
defines an ambient metric
\be
g_{MN}=\nabla_M X_N\, ,
\ee
where  $\nabla$ is its Levi-Civita covariant derivative.
It follows that the ambient metric is the double gradient of the homothetic
potential $\frac12 X^2$,
\be
g_{MN}=\frac 12 \nabla_M \partial_N X^2\, .
\ee
The zero locus of the potential defines the curved generalization of the null cone.
The requirement that $g_{MN}$ derive from a closed homothety along with an almost Ricci flat condition
determines a unique Fefferman--Graham ambient metric~\cite{FG} for a given conformal class of metrics.
Tractors are thus described in terms of ambient tensors via the equivalence relation
\be\label{equiv}
\Phi\sim\Phi + X^2 \, \chi\, ,
\ee
and the weight constraint
\begin{equation}
X\cdot \nabla \Phi = w\, \Phi\, .
\end{equation} 
The Thomas $D$-operator then has the ambient form
\be\label{Th}
D_M=2\, (X\cdot \nabla+d/2)\ \nabla_M-X_M\ \Delta\, ,
\ee
which is well defined on the above equivalence classes. Note that for a flat space, this is the generator of a conformal 
boost in  a momentum representation, which was the motivation for identifying the conformal cone as the space of lightlike states in~\cite{Gover:2009vc}.

Specializing to the  the case where $(\manM,[g_{\mu\nu}])$
%~$\manM$ 
is conformally flat,
the Fefferman--Graham metric is pseudo-Euclidean as in~\eqn{flatamb} and the homothety is generated by the vector field $Y\cdot\dl{Y}$.
A standard choice of gauge for the equivalence~\eqn{equiv} is the harmonic condition $\Delta\Phi=0$. 
So, functions on~$\manM$ are extended
off the submanifold by requiring
\begin{equation}
\label{cconst}
%  \dl{Y^M}\dl{Y_M}f=0\ \mbox{ and }\  {Y^M}\dl{Y^M}f=wf\,.
\dl{Y}\cdot \dl{Y}f=0\ \quad \mbox{ and }\ \quad  {Y}\cdot \dl{Y}\,f=wf\,.
\end{equation}
To see that this is possible one first extends the function to the entire cone using the second condition and then imposes the second. 
Generically, away from isolated values of the weight $w$ discussed later, this is straightforward.

The space singled out by~\eqref{cconst} is isomorphic to the space of functions on~$M$. Under this isomorphism, we see from~\eqn{Th} that
the differential operator~$\dl{Y^M}$ is mapped (up to a constant, but weight dependent,  coefficient) to the Thomas~$D$ operator. Taking into account~\eqref{cconst} this gives~$D^M \propto \frac{\partial}{\partial Y_M}$. Hence in our formula (since we work at a definite weight) we can simply replace~$D^M \to \frac{\partial}{\partial Y^M}$.

\begin{figure}
 \begin{center}
 \includegraphics[height=5cm,width=8cm]{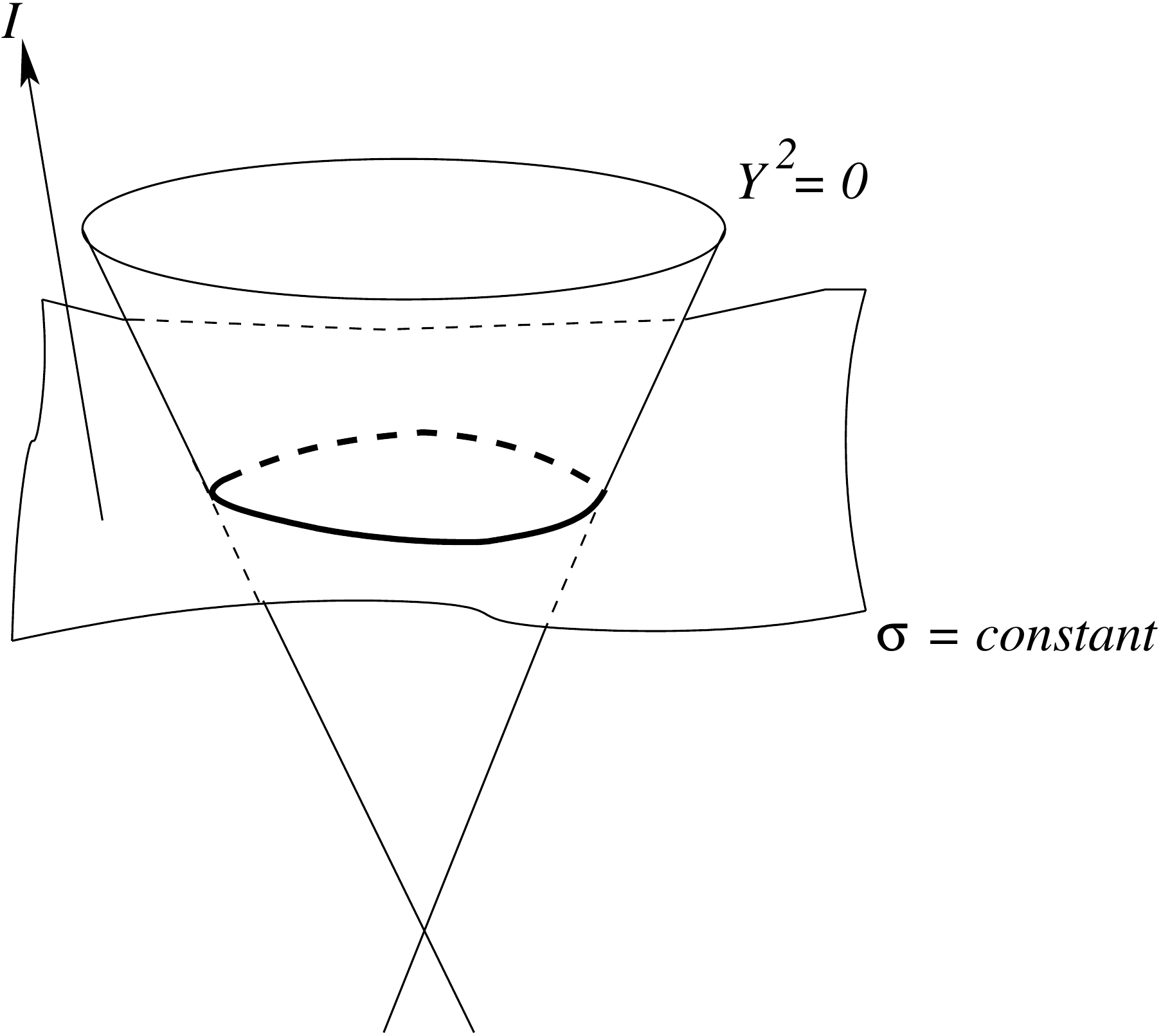}
 \end{center}
 \caption{\label{sigma} The ambient space description of a conformal manifold.}
 \end{figure}

Now, making a choice of scale by choosing a harmonic weight one function~$\sigma$ on the ambient space  
specifies a metric from the conformal class. For example the standard AdS, Minkowski and dS choices are
achieved via $\sigma$ $=$ $Y^d$, $Y^+$, $Y^0$, respectively. The submanifold is then the intersection of
the constant~$\sigma$ hypersurface with the cone $Y^2=0$ and inherits the metric obtained from the ambient one by pullback. 
Since $\sigma$ is harmonic, the scale tractor~$I_M=\frac1d D_M\sigma\sim \frac{\partial\sigma}{\partial Y^M}$ is the vector normal to the constant sigma surface.
This is depicted in Figure~\ref{sigma}.

This ambient space description allows us to enlarge the constraint algebra~~$\frak{g}$ in Figure~\ref{tractor-const}
to the following first class algebra
\begin{equation}
\label{aconst}
 \begin{gathered}
   \Delta=\dl{Y}\cdot\dl{Y} \,,\qquad h=(Y\cdot  \dl{Y}-w)\, ,\\[2mm]
\div=\dl{Y}\cdot \dl{Z}\,, \qquad \grad=Z\cdot \dl{Y}\,, \qquad \tr=\dl{Z}\cdot\dl{Z}\,,\\[2mm]
\IcdotD=I\cdot \dl{Y}\,, \qquad \tilde\div=I\cdot \dl{Z} \,,
 \end{gathered}
\end{equation}
The constraints in the first line are just those in~\ref{cconst}. Together with
the  constraints in the second line, these give the usual set of constraints
needed for a first quantized description of  massless higher spin fields in
$(d+2)$-dimensions, or constant $(d+1)$-dimensional constant curvature higher
spins taking into account the second, weight, constraint. On the other hand,
together, the  second and third lines are the tractor constraints of
Figure~\ref{tractor-const} expressed ambiently. Finally, the {\it r\^ole} of
those on the third line is to eliminate coordinates along the~$I$ direction. 

Now let us briefly discuss the special values of the weight~$w$. First observe that the factor~$X\cdot\nabla+d/2$ appearing in~\eqn{Th} implies the appearance of 
special eigenvalues of the operator $X\cdot \nabla$, or in other words special tractor weights. When the operator
$D_M$ acts on a field $\Phi$ of weight $w$, the interesting weight is at $w=1-\frac d2$,  while when acting on gauge parameters (which have 
weight $w+1$)
the special weight is at the value $w=-\frac d2$. Both these cases have been analyzed in~\cite{Gover:2008pt,Gover:2008sw}.
The first case, $w=1-\frac d2$, is the higher spin analog of the the canonical weight at which a conformally improved
scalar field enjoys conformal symmetry. Note that in dimension $d=4$, $w=1-\frac d2 = -1$. This value of the weight 
corresponds to maximally depth partially massless theories which are known to be conformally invariant~\cite{Deser:2004ji}.
The $s=1$ vector case of this series, is the usual statement that four-dimensional Maxwell theory is conformally invariant. 
In other dimensions, the value $w=1-\frac d2$ also corresponds to conformally invariant higher spin theories, but now
of the novel variety whose spin~1 progenitor was first introduced by Deser and Nepomechie in~\cite{Deser:1983tm,Deser:1983mm}.
The value $w=-\frac d2$ is apparently less interesting, since (for lower spins at least), it has been shown  at the level of
tractors, to amount simply to massive theories in which the relationships between various tractor components
of fields, and St\"uckelberg auxiliary fields are shuffled among one another.

Finally, we should also mention the
other origin of special weights, namely logarithmic obstructions to formally extending AdS boundary data into the bulk.
A detailed analysis of this phenomenon from a tractor view point has recently been given in~\cite{Gover:2011rz}. These appear at weights
such that $d+2w=2,4,6,\ldots$ and correspond to the special eigenvalues encountered in the conformal scattering 
study of~\cite{GZ}. In Section~\bref{sec:parent-tractor} we show how this formal extension can be analyzed within the
parent BRST formulation of tractor fields. In more physical terms these correspond to the holographic anomalies developed in detail in~\cite{Henningson:1998gx,deHaro:2000xn,Skenderis:2002wp}.

\section{Massive Higher Spin Fields}\label{radial}

In this Section we give three equivalent descriptions of the equations of motion of massive, $d$-dimensional, constant curvature, higher spin theories and their massless and partially massless limits. The first and simplest is in terms of $d$-dimensional, on-shell equations
in a constant curvature background. The second is in terms of massless theories in a $(d+1)$-dimensional flat background. The third, 
is a $(d+2)$-dimensional approach, where the conformally flat, constant curvature spacetime is realized as a slice of the light cone in
an ambient space with an additional time and space coordinate. The latter approach melds best with the tractor and BRST descriptions of
higher spins.

\subsection{On-shell Equations of Motion}

An ``on-shell'' description of massive higher spins is very simply given. In terms of a spin $s$ symmetric tensor $\phi(x,z)$
in the notation of Section~\bref{OSC} (so that ${\rm N}\, \phi = s\, \phi$) we 
impose\footnote{The dependence of the mass-like term on the spin was found in~\cite{Fronsdal:1978vb}
for $d=4$ and extended to generic dimensions (and the mixed-symmetry case) in~\cite{Metsaev:1997nj}.}
\be
\big(\nabla^2  +s - (s-2)(s+d-3)  \big)\, \phi = m^2 \phi \, ,\qquad {\rm div}\ \phi = {\rm tr}\ \phi =0\, .\label{onshell}
\ee
In this formula $\nabla^2=g^{\mu\nu}\nabla_\mu\nabla_\nu$ (with $\nabla_\mu$  the Levi-Civita connection) is the Bochner Laplacian and 
we  work in units 
\begin{equation}
\label{units}
\frac{2\scalebox{.9}{\Rho}}{d} = \frac{2\Lambda}{(d-1)(d-2)}=-1\, .
\end{equation} 
This assumes a negative scalar curvature, and therefore an AdS background. The cosmological constant $\Lambda$ can be reinstated
in any formula by a dimensional analysis, thereafter extending its validity to either sign of $\Lambda$. Notice that the mass squared is defined as the eigenvalue of the Laplacian up to an overall constant shift so that
$m^2=0$ corresponds to the standard massless limit. 

In AdS space, there is a global timelike Killing vector whose eigenvalue can be used to define energies. The lowest energy eigenvalue $E_0$ can be used as an alternate definition of mass 
via~\cite{Metsaev:2003cu}\footnote{For $s=1$ and $s=2$, this relation was found in \cite{Mueck:1998iz} and \cite{Polishchuk:1999nh} respectively.}

\begin{equation}\label{E0}
E_0=\frac{d-1}{2}+\sqrt{m^2+\Big(s-2+\frac{d-1}{2}\Big)^2}\,.
\end{equation} 
It would be desirable to have a definition of the mass that did not depend on underlying isometries but rather the  underlying
geometry of the model. As we shall see, this is provided by the weights of tractors.

For generic values of~$m$,
and constant curvature backgrounds, the above equations can be derived from a constraint analysis of a set of equations
of motion coming from an action principle. For example, when $s=2$, we have
\begin{equation}
\big(G_{\rm E}+G_{\rm PF}\big)\, \phi =0\, ,     \nonumber                                         
\end{equation}
\bea
&G_{\rm E} = \square  - 2d+2 - {\rm grad}\, {\rm div}+\frac12 \big[{\rm grad}^2\, {\rm tr} +{\rm g} \, {\rm div}^2\big]
-\frac12 {\rm g} \big[\square - d+1\big]{\rm tr}
\, ,&\nn\\[3mm]
&G_{\rm PF} = -m^2 \big(1-\frac12\, {\rm g}\, {\rm tr}\big)\, ,&
\eea
which follows from the action  $S=\int \big(\phi,[G_{\rm E}+G_{\rm PF}]\phi\big)$. The self-adjoint operator $G_E$ is often called
an Einstein operator since $G_E\, \phi$ is the linearized Einstein tensor. The second term in the field equation is the standard Pauli--Fierz mass term. 

In the above formul\ae ~we have introduced the constant curvature Lichnerowicz wave operator
\begin{equation}\label{algebra}
\square = [{\rm div},{\rm grad}]-2 \big({\rm g}\, {\rm  tr} - {\rm N}({\rm N} + d -2)\big) =\nabla^2 -{\rm g}\, {\rm  tr} + {\rm N}({\rm N} + d - 2)\, .
\end{equation} 
It commutes with all the operators in~\eqn{ops}. 
By tuning the  value of the mass  we expect to find massless or partially massless theories. These can be detected in 
an on-shell approach by searching for residual gauge invariances of the equations~\eqn{onshell}. For example the condition ${\rm div}\, \phi=0$ is invariant under
\be
\delta \phi = {\rm grad}\,  \xi
\ee
if 
\begin{equation}
{\rm div}\, \xi = {\rm tr} \, \xi = \big(\nabla^2 - {\rm N}({\rm N}+d-2) \big)\xi=0\, .                                                                                       \end{equation} 
Thus, varying the remaining on-shell equations~\eqn{onshell} we learn
\be
m^2=0\, .
\ee
This value of the mass parameter corresponds to massless fields propagating on a constant curvature background.
Specializing to our spin two example, the gauge invariance $\delta \phi={\rm grad}\, \xi$ 
is precisely a linearized diffeomorphism so $\phi$ describes massless spin~2 excitations ({\it i.e.} linearized gravitons).
However, it is a well known fact~\cite{Deser:1983mm,Deser:1983tm} that when $m^2=-d+2$  a partially massless spin~2 theory with gauge invariance
\begin{equation}
\delta \phi = \big({\rm grad}^2-{\rm g} \big)\, \xi\, ,
\end{equation} 
with a scalar gauge parameter $\xi$, arises. When $\xi$ obeys the onshell condition $(\nabla^2-d)\xi=0$, this is indeed 
an invariance of the on-shell equations~\eqn{onshell}. Note the parameter~$m$ is only real for a dS background (for which the sign in the partially massless tuning flips). In fact, the representation of the constant curvature algebra is only unitarity in this case too~\cite{Deser:2001pe,Deser:2001us,Deser:2003gw}.

The above massless and partially massless theories are examples of a more general theory. In general the on-shell equations~\eqn{onshell} enjoy ``depth~$t$'', higher derivative on-shell gauge invariances of the form
\begin{equation}
\delta \phi = \big({\rm grad}^t + \cdots\big) \, \xi\, ,\qquad 1\leq t\leq s\, ,
\end{equation} when the mass  obeys
\be\label{mu2}
m^2=  -(t-1)(2s-t+d-4)\, .
\ee
This formula can be re-expressed as
\begin{equation}
m^2 =\Big(w+\frac{d-1}{2}\Big)^2-\Big(s-2+\frac{d-1}{2}\Big)^2\, , \mbox{ with } w=s-t-1\, .
\end{equation}
The parameter~$w$ will play an important {\it r\^ole} in the remainder of this Article, as the weight of tractors or the eigenvalue
of the Euler vector field in a radial Scherk--Schwarz reduction. A review of the description of massive higher spins on constant curvature backgrounds as the log radial reduction of massless theories on a flat background of one higher 
dimension~\cite{Biswas:2002nk,Hallowell:2005np} is our next topic.

\subsection{Log Radial Reduction}
\label{log R}

\begin{figure}
 \begin{center}
 \includegraphics[scale=.5]{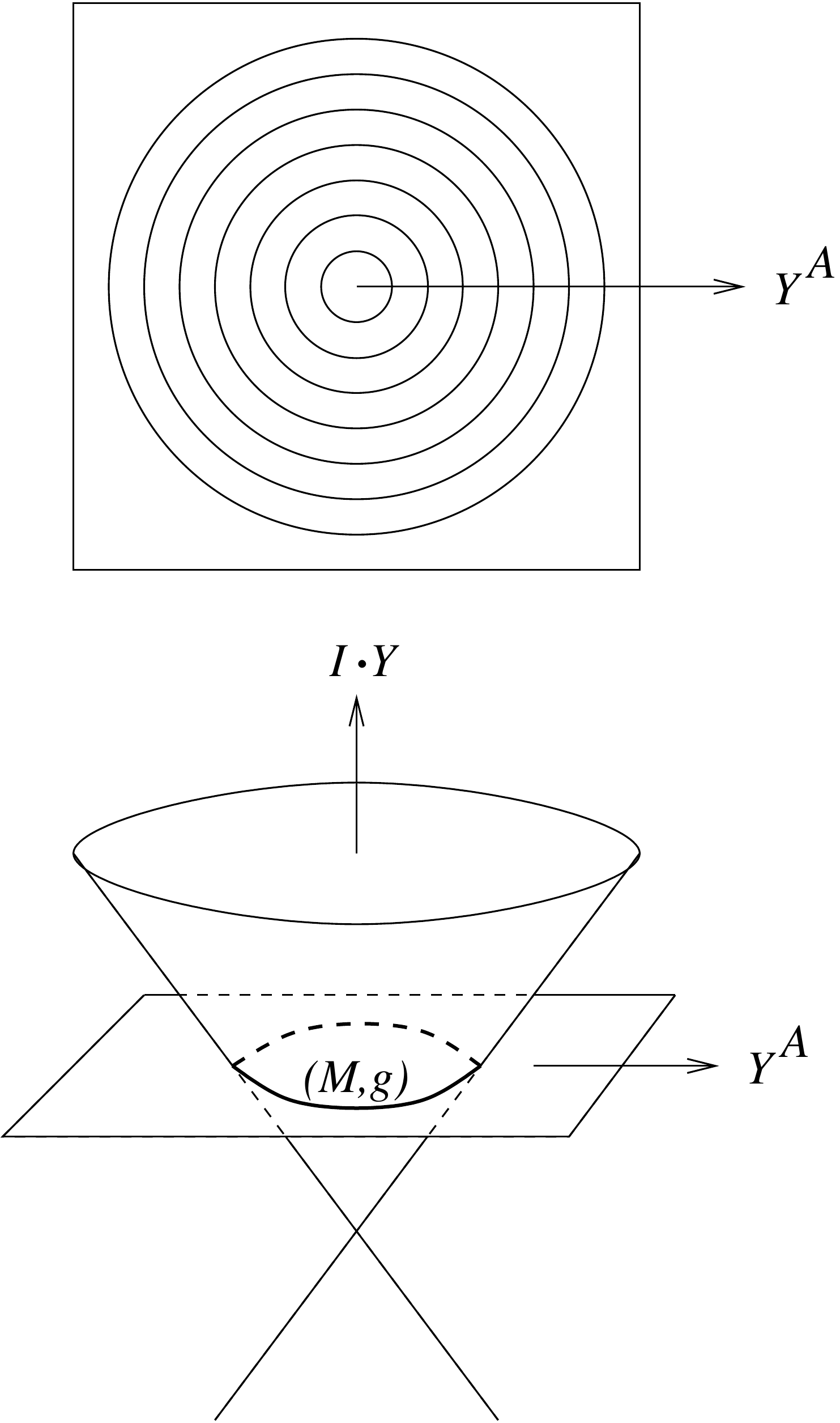}
 \end{center}
 \caption{Projective model for conformal geometry depicted in Euclidean signature.\label{cone}}
 \end{figure}

It has long been known, since the work of Scherk and Schwarz~\cite{Scherk:1979zr}, that massive fields can be obtained as the dimensional 
reduction of massless ones in one higher dimension. Biswas and Siegel~\cite{Biswas:2002nk} sharpened this proposal by suggesting that massive fields 
on constant curvature spaces could be obtained by dimensional reduction of a flat space theory in one higher dimension along a conformal, radial, isometry. Their suggestion was verified for all higher spins in~\cite{Hallowell:2005np}. The Riemannian signature 
version of this reduction is sketched in the top half of Figure~\ref{cone}.

We call the $(d+1)$-dimensional flat space analogs of the geometric operators in~\eqn{ops}
\begin{equation}
\begin{gathered}
 \label{d+1ops}
  {\bf g}=Z\, . \,  Z\, ,\qquad {\bf N}=Z\, . \,  \frac{\partial}{\partial Z}\, ,\qquad {\bf tr}=\dl{Z}\, . \, \dl{Z}\, ,\\ 
 {\bf grad}=Z \, . \, \frac{\partial}{\partial Y}\, ,\qquad {\bf div} = \dl{Z}\, . \, \dl{Y}\,.
\end{gathered}
\end{equation}
Here and below we use $F\, . \,  G=F^AG_A$ to denote the contraction between a pair of $(d+1)$-dimensional indices.
The Curtwright--Fronsdal equations of motion for massless higher spins~\cite{Fronsdal:1978rb,Curtright:1979uz} are then given in terms of a rank $s$ symmetric
tensor (so that ${\bf N}\, \varphi(Y,Z)=s\varphi(Y,Z)$)
\begin{equation}\label{CF}
\big({\bf \Delta}-{\bf grad}\, {\bf div} + \frac 12 \, {\bf grad}^2 \, {\bf tr} \big)\,  \varphi = 0 = {\bf tr}^2 \varphi \, ,
\end{equation} 
where ${\bf \Delta}=[{\bf div}\, , \, {\bf grad}]=\dl{Y}\, . \,  \dl{Y}$.   

The main idea of the radial reduction is that the $(d+1)$-dimensional field content and gauge invariances give the St\"uckelberg
fields and local shift symmetries, respectively, in the $d$-dimensional massive, reduced theory. This can be easily seen by examining the
gauge invariance of the Curtwright--Fronsdal equations
\be\label{mg}
\delta \varphi = {\bf grad} \, \xi\, ,\qquad {\bf tr}\, \xi=0\, ,
\ee
in the coordinate system for the $(d+1)$-dimensional metric\footnote{To obtain dS slices the metric on the space $Y^A$ has signature
$(1,d)$ and the sign of $du^2$ in this formula flips.}
\be
ds^2=dY^A dY_A = \exp(2u)\, \Big(-du^2 + ds^2_{\rm AdS}(x)\Big)\, ,\label{radialds2}
\ee
where $u$ is the log of the radial coordinate. In this coordinate system, the radial conformal isometry
is \be \label{euler}N_Y\equiv Y\, . \, \frac{\partial}{\partial Y}=\frac{\partial}{\partial u}\, .\ee
For the simplest case where $\varphi$ is a scalar field, the Scherk--Schwarz reduction

\begin{equation}
\frac{\partial}{\partial u} = w\, ,
\end{equation} 
solved via
$
\varphi(Y) = e^{wu}\,  \phi(x)
$,
converts the massless, $(d+1)$-dimensional flat wave equation ${\bf \Delta}\varphi=0$ to the massive $d$-dimensional constant curvature one
\begin{equation}
\nabla^2\phi = 
%-\frac{2\Lambda}{(d-1)(d-2)}
\Big[\Big(w+\frac{d-1}{2}\Big)^2-\Big(\frac{d-1}{2}\Big)^2\, \Big]\, \phi \, .
\end{equation} 
The relationship between the mass term and the parameter $w$ is the same as that of the mass--Weyl weight relationship 
of the tractor approach given in~\cite{Gover:2008pt,Gover:2008sw}. 

For the full log radial reduction of the Curtwright--Fronsdal higher spin equations~\eqn{CF}, we refer to the literature, but by focusing
on the gauge invariance~\eqn{mg} the key structure can be uncovered. For concreteness consider a spin~2 field $\varphi(Y,Z)$ which 
we expand in terms of rank two, one and zero,  $d$-dimensional fields as
\begin{equation}
\varphi = e^{wu}\,\big(h+A \, Z^{(u)} + \chi \big[Z^{(u)}\big]^2\big)\, .
\end{equation} 
Then\footnote{To verify this computation, some geometric data is useful: The flat metric~\eqn{radialds2} equals $ds^2= -\big(E^{(u)}\big)^2+E^mE_m$ with vielbeine $(E^{(u)},E^m)=e^u(du,e^m)$ where $e^m$ are the AdS vielbeine. In these coordinates the spin connection is $\Omega^{(u)n}=e^n$ and 
$\Omega^{mn}=\omega^{mn}$ with $de^m + \omega^m{}_n \wedge e^n=0$. Thus the covariant derivative acting on symmetric forms can be represented by 
$\boldsymbol{\nabla}=\partial+(Z^m\omega_m{}^n -Z^{(u)} e^n)\, \frac{\partial}{\partial Z^n} -e^nZ_n\frac{\partial}{\partial Z^{(u)}}$. The result~\eqn{grad2grad} follows accordingly.
}
expanding ${\bf grad}$ in powers of $Z^{(u)}$
\be\label{grad2grad}
 {\bf grad}=e^{-u}\Big\{{\rm grad} - {\rm g}\, \frac{\partial}{\partial Z^{(u)}}+Z^{(u)}\Big(\frac{\partial}{\partial u} - {\rm N}\Big)\Big\}\, ,
\ee
we may easily compute the variations $\delta \varphi = {\bf grad}\,  \xi$ with $\xi=e^{(w+1)u} (\zeta+\alpha Z^{(u)})$ and find
\be
\begin{aligned}
\label{stg}
\delta h &={\rm grad}\,  \zeta - {\rm g} \, \alpha\,,\\[2mm]
\delta A &= w  \zeta + {\rm grad}\, \alpha \,,\\[2mm]
\delta \chi &= (w+1)\,  \alpha\,.
\end{aligned}
\ee
These equations show that the vector and scalar fields $(A,\chi)$ are
St\"uckelberg auxiliary fields when $w\neq 0,1$. At $w=0$, the scalar $\chi$ is
still an auxiliary field and may be gauged away, while the vector $A$ is
invariant under the remaining $\zeta$-transformations. It may therefore
\hypertarget{decouples}{consistently be set to zero and anyway decouples from equations of
motion}. The remaining field
$h$ then transforms as a linearized metric tensor under diffeomorphisms, and
therefore describes massless spin~2 excitations. At $w=-1$, the auxiliary $\chi$
is inert and can be discarded. The vector $A$ is auxiliary and can be gauged
away, but thereafter, all further $\alpha$-transformations must be accompanied
by a compensating transformation $\zeta={\rm grad}\, \alpha$ so that
\begin{equation}
\delta h = \big({\rm grad}^2-{\rm g}  \big)\alpha\, .
\end{equation} 
This is a depth $t=2$ partially massless gauge transformation, so the $w=-1$ theory describes partially massless spin~2.

In summary, therefore, we learn that massive constant curvature higher spins are described by the $(d+1)$-dimensional 
Curtwright--Fronsdal equations plus a Scherk--Schwarz log radial reduction condition
\begin{equation}
\begin{gathered}
({\bf \Delta}-{\bf grad}\, {\bf div} + \frac 12 \, {\bf grad}^2 \, {\bf tr} ) \, \varphi = 0 = {\bf tr}^2 \varphi \, ,\\
\label{hslog R}
Y\, .\,  \frac{\partial}{\partial Y} \varphi = w\varphi\, ,
\end{gathered}
\end{equation} 
with mass
\be\label{mass}
m^2 = 
\Big(w+\frac{d-1}{2}\Big)^2-\Big( s-2+\frac{d-1}{2} \Big)^2\, .
\ee
Standard massless and partially massless models appear at special weights $w=s-2$ 
and $w=-1,\ldots, s-3$, respectively~\cite{Gover:2008pt,Gover:2008sw}.  For the  case of AdS space, let us also comment on the relationship between the energy $E_0$ and the weight $w$.
Rewriting  the mass parameter in terms of $w$ and $E_0$ using~\eqref{E0} one finds:
\begin{equation}
 E_0(E_0-d+1)=w(w+d-1)\,.
\end{equation} 
Two solutions are possible $E_0=-w$ and $E_0=w+d-1$.  The second value is selected
on unitarity grounds (see {\it e.g.} the discussion in~\cite{Metsaev:1997nj}).

The last description of massive fields we give is in terms of a $(d+2)$-dimen\-sional ambient space.
Let us preempt that discussion by noting that the gauge transformations~\eqn{stg} also follow directly from the
tractor formulation. In fact, the tractor gauge transformations $\delta \Phi^{MN}=D^{(M}\Phi^{N)}$,
written out for the independent scalar, vector and tensor components of the tractor tensor~$\Phi^{MN}$
lead exactly to the equations~\eqn{stg}. Coupled with the discussion of the ambient space construction of the tractor calculus
in Section~\bref{ambient space}, this strongly suggests a $(d+2)$-dimensional formulation of massive higher spin theories, which we now present.

\subsection{Ambient Approach}\label{sec:amb-appr}

In Section~\bref{ambient space}  we described the ambient approach to conformal geometry, and in particular how to select a Riemannian manifold from a conformal class of such manifolds by studying the intersection of the constant~$\sigma$ surface with the
ambient null cone (see Figure~\ref{sigma}). We now want to relate this description with the $(d+1)$-dimensional log radial one
in the previous Section. Pictorially, the idea is given in Figure~\ref{cone}. The ambient scalar field $\sigma(Y^M)$ determines the hypersurface on which one subsequently reduces log radially.

Focussing on the conformally flat setting with ambient metric~\eqn{flatamb},
the choice
\begin{equation}
\sigma=\frac{Y^++Y^-}{\sqrt{2}}:=Y^{{}^{\sss (d+1)}}
\end{equation} 
yields AdS space at the intersection of hypersurfaces $\sigma=1$ and $Y^2=0$. 

On the other hand, if we first impose the choice $\sigma=1$ before the homogeneity and 
cone conditions, then we obtain a flat hypersurface with  accompanying signature $(d-1,2)$ metric; 
\begin{equation}
ds^2_{\rm flat}= dY^A dY_A\, ,
\end{equation} 
which is exactly the $(d+1)$-dimensional flat space of the log radial reduction in the previous Section.

Our aim is to relate tractor equations to the massless equations of the previous Section, on the hypersurface $\sigma=1$ along with a log radial reduction condition controlling the $d$-dimensional masses.
The way this works is  very simple; consider as an example, a
massive scalar field, which, according to~\cite{Gover:2008pt,Gover:2008sw}, is described by a weight~$w$ tractor $\Phi$ subject to 
\begin{equation}
\IcdotD \, \Phi = 0\, .
\end{equation}
 The scale tractor $I^M$ is given by~\eqn{scale tractor}, and can be explicitly, ambiently computed to yield a  timelike vector. It is then convenient to redefine $Y^{d+1},Y^d$
by an orthogonal transformation in order to achieve
\begin{equation}
I\cdot \frac{\partial}{\partial Y} = \frac{\partial}{\partial Y^{{}^{\sss (d+1)}}}\, .
\end{equation} 
Note that $I\cdot \frac{\partial}{\partial Y}$ is perpendicular to our  hypersurface.

Picking a representative for the cone equivalence relation in~\eqn{equiv} by the harmonic gauge choice~\eqn{cconst}
allows us to identify the Thomas $D$-operator with the ambient  gradient operator so that $\IcdotD\sim \frac{\partial}{\partial Y^{{}^{\sss (d+1)}}}$.
Hence the  tractor equation of motion performs the reduction to the flat hypersurface; $\Phi(Y^{{}^{\sss (d+1)}},Y^A)=\varphi(Y^A)$. 
Therefore the field $\varphi(Y^A)$ is subject to 
\be\label{scalar}
{\bf \Delta}\,  \varphi = 0 =\Big(Y . \frac{\partial}{\partial Y}-w\Big)\, \varphi\, ,
\ee
which are exactly the equations for a massless, log radially reduced scalar field described in Section~\bref{log R}.

Let us now repeat the above scalar computation for the case of the tractor higher spin equations in Figure~\ref{tequations};
our aim is to show that they are equivalent to the equations~\eqn{hslog R}.
To start with, we express the tractor operators of Section~\bref{OSC} in terms of the $(d+1)$-dimensional operators in~\eqn{d+1ops} 
by replacing $D^M \to \frac{\partial}{\partial Y^M}$ and treating $Z^{{}^{\sss (d+1)}}$ and $Y^{{}^{\sss (d+1)}}$ as auxiliary variables
 \begin{equation}
\frak{Grad}=Z^{{}^{\sss (d+1)}}\frac{\partial}{\partial Y^{{}^{\sss (d+1)}}}+{\bf grad}\, ,\qquad
\IcdotD = \frac{\partial}{\partial Y^{{}^{\sss (d+1)}}}\, ,\qquad
\wt{\frak{Div}}=\frac{\partial}{\partial Z^{{}^{\sss (d+1)}}}\, ,\nonumber
\end{equation} 
\be
\frak{Div}=\frac{\partial}{\partial Y^{{}^{\sss (d+1)}}}\frac{\partial}{\partial Z^{{}^{\sss (d+1)}}}+{\bf div}\, ,\qquad
\frak{Tr}=\Big(\frac{\partial}{\partial Z^{{}^{\sss (d+1)}}}\Big)^{\! 2}+{\bf tr}\, .
\ee
We also need the ambient  Laplacian 
 \begin{equation}
\frac{\partial}{\partial Y}\cdot\frac{\partial}{\partial Y}=\Big(\frac{\partial}{\partial Y^{{}^{\sss (d+1)}}}\Big)^{\! 2}\ +\ {\bf \Delta}\, .
  \end{equation} 

Applying these results to the gauge invariances of the tractor equations
\begin{equation}
\delta \Phi = Z^{{}^{\sss (d+1)}}\frac{\partial\varepsilon}{\partial Y^{{}^{\sss (d+1)}}}+{\bf grad}\, \varepsilon\,   ,\qquad \frac{\partial^2\varepsilon}{(\partial Z^{{}^{\sss (d+1)}})^2}+{\bf tr}\, \varepsilon=0=\frac{\partial\varepsilon}{\partial Z^{{}^{\sss (d+1)}}}\, ,
\end{equation} 
we firstly learn that the gauge parameter $\varepsilon$ is $Z^{{}^{\sss (d+1)}}$-independent and trace-free
\begin{equation}
\varepsilon = \varepsilon(Y^{{}^{\sss (d+1)}},Y^A)\, ,\qquad {\bf tr}\, \varepsilon = 0\, .
\end{equation} 
Expanding the field~$\Phi$ in powers of $Z^{{}^{\sss (d+1)}}$
\be
\Phi = \varphi(Y^M,Z^A) + Z^{{}^{\sss (d+1)}} \, \chi(Y^M,Z^A) + {\cal O}\big((Z^{{}^{\sss (d+1)}})^2\big)\, ,
\ee
it follows that
\be
\delta \varphi = {\bf grad} \, \varepsilon \, ,\qquad \delta \chi = \frac{\partial \varepsilon}{\partial Y^{{}^{\sss (d+1)}}}\, .
\ee
Observe, that we already expect the $Y^{{}^{\sss (d+1)}}$ independent piece of $\phi$ to be the physical field content since
it has the correct gauge transformation. Analyzing the triplet of tractor equations 
$\wt{\frak{Div}}{}^{\, ^{\scriptstyle 2}} \Phi = \frak{Tr}\ \wt{\frak{Div}} \, \Phi =  \frak{Tr}^{\, 2}\Phi  = 0$ we quickly find
\begin{equation}
\Phi =\varphi + Z^{{}^{\sss (d+1)}} \, \chi\, , \quad {\bf tr}\, \chi = {\bf tr}^2 \varphi = 0\, .
\end{equation} 
Then the final pair of tractor equations $(\IcdotD - \frak{Grad}\, \wt{\frak{Div}})\Phi$ $=$ $\big(\frak{Div}\, \Phi - \frac12 \, \frak{Grad} \frak{Tr}\big) \Phi$ $=0$ give
\bea
\frac{\partial \varphi}{\partial Y^{{}^{\sss (d+1)}}}  &=& {\bf grad}\, \chi\, ,\nn\\[3mm]
\frac{\partial\chi}{\partial Y^{{}^{\sss (d+1)}}}  &=& \Big[\frac12\, {\bf grad} \, {\bf tr}-{\bf div}\Big]\varphi\, .
\eea
Therefore, from the harmonic condition~$\frac{\partial}{\partial Y^N}\frac{\partial}{\partial Y_N}\varphi = 0$ we obtain
\begin{multline}
{\bf \Delta} \phi ~~=~~ -\Big(\frac{\partial}{\partial Y^{{}^{\sss (d+1)}}}\Big)^2\, \varphi ~~=~~ -\frac{\partial}{\partial Y^{{}^{\sss (d+1)}}}\, {\bf grad} \, \chi ~~= \\
= ~~{\bf grad}\,   \big[{\bf div} -\frac12 \, {\bf grad}\,  {\bf tr}\big]\, \varphi\, .                                                                                                                                                                                                                                                    \end{multline}
Which exactly matches the first equation of those we were aiming for in~\eqn{hslog R}.

Our computation is not quite complete, because we must still eliminate the field $\chi$ along with all dependence on the ambient
coordinate $Y^{{}^{\sss (d+1)}}$. In fact, it is not difficult to check that expanding the fields $\varphi$ and $\chi$ as well as the gauge parameter~$\varepsilon$ in Taylor series about $Y^{{}^{\sss (d+1)}}=1$, only terms of order zero and one in $(Y^{{}^{\sss (d+1)}}-1)$ are independent once one imposes the ambient harmonic condition. In turn the order one part of the gauge parameter can then be used to algebraically remove the order zero part of $\chi$. All remaining fields save for the leading order term of $\varphi$ are then algebraically dependent via the field equations. Hence $\varphi=\varphi(Y^A)$, subject to exactly equations~\eqn{hslog R}. Therefore we now see that  the tractor equations displayed in Figure~\eqn{tequations}
describe massive higher spins. Let us now investigate their cohomological, BRST origins which yields an alternative, elegant, proof of the conjecture of~\cite{Gover:2008pt,Gover:2008sw}.

\section{BRST Description of Massive Higher Spins}

\label{BRST}

In this Section, after a brief review of first quantized BRST techniques, we use this approach to derive the massive tractor equations
presented in the Introduction in Figure~\ref{tequations}. We then give a very simple BRST constraint analysis demonstrating directly the equivalence of the tractor equations and the $d$-dimensional on-shell system~\eqn{onshell}. Finally we present a parent BRST approach which overlies the other BRST approaches. 

\subsection{BRST First Quantized Approach}\label{sec:BRST-general}

Let $g_a(x,\dl{x})$ be a set of differential operators defined on the space $\cal V$ of functions
on the space-time manifold $\manM$ taking values in some internal space. 
Suppose also that the set is compatible in the sense that $\commut{g_a}{g_b}$ is again proportional to $g_c$.
Moreover, to simplify the exposition we even assume that constraints $g_a$ form a Lie algebra~${\mathfrak g}$ so that
\begin{equation}
[g_a,g_b]=f_{ab}^c \, g_c\, ,
\end{equation} 
for some constants $f_{ab}^c$.
Let us also assume for simplicity that $g_a$ are independent so that there are no relations between them.

There can be two physical interpretations of the set of compatible constraints.
The first is to treat them as the first class ones of the quantum mechanical model
with functions from $\cal V$ playing the {\it r\^ole} of wave functions. The second, which we
focus on here, is to relate them to  equations of motion and
generators of gauge symmetries of a local gauge field theory whose 
fields are defined on $\manM$ and take values in the internal space. In order to distinguish between the equations of
motion and generators of gauge symmetry one needs to pick a polarization that
identifies which of  the initial set of constraints correspond to genuine
equations  of motion as opposed to gauge symmetry generators.

To define and study this gauge field theory it is useful to employ  BRST techniques. To this end, to each constraint $g_a$ one associates a pair of ghost variables
$(c^a,b_a)$ that are fermionic (in general the parity of $c^a,b_a$ is opposite to that of $g_a$) and satisfy the following
commutation relations and ghost number assignments
\begin{equation}
 \commut{c^a}{b_b}=\delta^a_b\,.\qquad \gh{c^a}=1\,,\quad \gh{b_a}=-1\,.
\end{equation} 
One then builds the fermionic BRST operator
\begin{equation}
 \Omega=c^a g_a-\half c^ac^bf_{ab}^c b_c\,,\qquad \gh{\Omega}=1\, .
\end{equation} 
Its nilpotency, $$\Omega^2=0\, ,$$ is ensured by compatibility of the constraints. 

The polarization of the constraints can be implemented by the choice of
representation of the ghosts: Variables $(c^a,b_a)$ that are
associated to genuine equations of motion are represented in a coordinate
representation, {\it i.e.},  $c^a$ acts by multiplication while $b_a=\dl{c^a}$.
Ghosts corresponding to generators of gauge symmetries are represented in
a momentum representation, {\it i.e.}, $c^a=\dl{b_a}$. The internal space is then
extended to a ``BRST Hilbert space'' $\cH$ by polynomials of the coordinate ghosts/anti-ghosts 
 to form a representation of the  entire operator algebra.
Through the ghost variables,  the representation space acquires a grading by 
ghost degree $\cH=\oplus_i \cH^i$ whith $\gh{\cH^i}=i$. We use the convention
that, in this representation, the ghost degree is normalized so that $\gh{1}=0$.

The physical fields (along with all auxiliary and St\"uckelberg fields) are by
definition contained in the set of  ghost number zero elements of the representation
space, {\it i.e.}, they take values in $\cH^0$. If $\Psi(x)$ is a general $\cH^0$-valued field then the equations of motion are compactly written as
\begin{equation}
 \Omega \, \Psi(x)=0\, .
\end{equation} 
Gauge parameters are identified with $\cH^{-1}$-valued functions~$\Xi(x)$. The gauge
transformation determined by these is 
\begin{equation}
\label{brst-gauge}
 \delta_\Xi \Psi=\Omega \, \Xi\, .
\end{equation} 
That this is a symmetry of the equations of motion $\Omega\, \Psi=0$ follows immediately
from nilpotency of~$\Omega$. Fields with different ghost degrees also have a natural
interpretation as reducibility relations ({\it alias} gauge for gauge symmetries), Bianchi identities {\it etcetera}. An exhaustive
treatment of these interpretations can be given in terms of the BRST formulation of the resulting gauge theory
(see {\it e.g.}~\cite{Barnich:2003wj,Barnich:2004cr}).

In many cases, all representations of the ghost commutation relations are
equivalent. For instance, if ghosts $c,b$ are fermionic then it does not matter
which representation is chosen. The only consequence of representing $c,b$ as
$c=\dl{b}, b=b$ instead of $b=\dl{c}, c=c$ is an overall shift of the ghost
number. The isomorphism between these two representations sends $b$ to $1$ and 
$1$ to~$c$. In other words, if all ghosts are fermionic and $l$ of them
are assigned to gauge generators, then instead of representing them in momenta
representation one can equivalently require physical fields to appear at ghost
degree $l$ and the gauge parameters at ghost degree $l-1$.
In this case, the equations of motion and gauge symmetries encoded in the BRST
operator can be seen,  respectively, as the cocycle and the coboundary conditions
for $l$-th cohomology group of the Lie algebra of the constraints with
coefficients in the space $\cal V$ of internal space valued functions. Note that if all
ghosts are represented in the coordinate representation then the ghost degree is
just the usual degree on the Lie algebra complex.

As an example, consider explicitly the usual case where only one gauge generator is present. An example of this
is the case of totally symmetric higher-spin fields. In such cases, there is a very simple trick to characterize the Lie algebra cohomology $H^1({\frak{g}},{\cal V})$.
 One writes out
the Lie bracket as a commutator (since we have a representation on ${\cal V}$) 
\be\label{commutator}
g_a g_b - g_b g_a = f_{ab}^c g_c\, ,
\ee
and then replaces the rightmost generator in each term of this equation with a ${\cal V}$-valued field labeled by that generator
\be
g_a \phi_b - g_b \phi_a = f_{ab}^c \phi_c\label{eom}\, .
\ee
This set of~\raisebox{1mm}{\scalebox{.8}{$\left(\!\!\!\begin{array}{c}\dim \frak g \\[1mm] 2 \end{array}\!\!\!\right)$}} relations are then the equations of motion
which, by virtue of~\eqref{commutator}, obviously enjoy the gauge invariance 
\be
\delta \phi_a = g_a \varepsilon\, ,\label{gauge}
\ee
for any ${\cal V}$-valued gauge parameter~$\varepsilon$.

In general, some of the relations in~\eqn{eom} will allow a subset of fields to
be eliminated algebraically  because the structure constants on the right hand
side allow some of the fields to be expressed in terms of others. Moreover, if
some of the operators $g_a$ appearing in the gauge transformation~\eqn{gauge}
can be inverted algebraically, then the corresponding field is a
St\"uckelberg auxiliary field and can be gauged to zero. In general, for higher
spin theories, these methods leave a ``minimal'' covariant field content subject
to off-shell equations of motion.

More generally, one can show~\cite{\BGST} that if the BRST Hilbert space~$\cH$  can be decomposed as 
$$\cH=
\cE\oplus\cF\oplus \cG$$ 
such that the only solution to 
$$
(\Omega f)|_{\cG}=0\, , \qquad f\in {\cF}\, ,
$$
(here ``${}|_{\cG}$'' denotes projection onto $\cG$) is $f=0$, then 
 all fields associated to
$\cF$ and~$\cG$ are in fact generalized auxiliary fields (these are standard auxiliary fields and St\"uckelberg fields along with
their associated ghost fields and antifields). This implies that the system can be reduced to $\cE$-valued fields. The reduced system is again a  BRST first-quantized model but now  with $\cE$ replacing $\cH$ and the nilpotent operator $\Omega_{\red}$
replacing $\Omega$ (see~\cite{\BGST} for the explicit structure of $\Omega_{\red}$ and further details).  

In general, the spaces $\cF$ and $\cG$ can be identified using  homological arguments (although in simple cases 
a brute force computation suffices). In particular, if the space~$\cH$  is graded 
such that  the degree is bounded below, and the lowest degree term $\Omega_{-1}$ in the BRST charge $\Omega$ acts algebraically,  then one takes $\cG={\rm Im}(\Omega_{-1})$ and chooses $\cF$ and $\cE$ such that ${\rm Ker}(\Omega_{-1})=\cG\oplus\cE$ (where $\cE$ is isomorphic to the $\Omega_{-1}$-cohomology) and $\cF$ is a complementary
subspace. In other words, contractible pairs for $\Omega_{-1}$ are generalized auxiliary fields for 
the entire BRST operator. The reduced BRST operator $\Omega_{\red}$ acting on $\cE$-valued functions can  then be constructed iteratively order by order in grading. A detailed description  can be found in~\cite{\BGST,\BGadS}.

In the next Section we explicitly apply the above procedure to the constraint 
algebra~\eqn{tractor-const} and show that it yields exactly the conjectured
tractor equations given in Figure~\ref{tequations} and discussed in the
Introduction.

\subsection{Tractor Equations of Motion}
\label{Brute}
We now consider 
the first class constraint algebra
\be
\frak{g}=\{\frak{Grad},\IcdotD,\frak{Div},\wt{\frak{Div}},\frak{Tr}\}\, ,
\ee
of Figure~\eqref{tractor-const} acting on totally symmetric tractor tensors as given in~\eqref{calV}. 
Since~$\frak{g}$ is dimension five, the model  has five gauge fields 
\be
\Big\{\phi_{\frak{Grad}},\phi_{I\cdot D},\phi_{{\frak{Div}}},\phi_{\wt{\frak{Div}}},
\phi_\frak{Tr}\Big\} \, ,
\ee 
with gauge transformations as in~\eqn{gauge}
\begin{equation}
\begin{gathered}
\label{GAUGES}
\delta\phi_{\frak{Grad}}={\frak{Grad}}\, \varepsilon\, ,\qquad
\delta\phi_{I\cdot D}={\IcdotD}\, \varepsilon\, ,\qquad
\delta\phi_{{\frak{Div}}}={{\frak{Div}}}\, \varepsilon\, ,\\
%
%\vspace{-.2cm}
%
\delta\phi_{\wt{\frak{Div}}}={\wt{\frak{Div}}}\, \varepsilon\, ,\qquad
\delta\phi_\frak{Tr}=\frak{Tr}\, \varepsilon \, .
\end{gathered}
\end{equation} 
There are ten gauge invariant equations of motion\footnote{The eight most important of these are listed explicitly in equations~(\ref{t1}, \ref{t2}, \ref{t3}, \ref{t4}, \ref{t5}, \ref{t6}, \ref{t7}, \ref{t8}).} of the form~\eqref{eom}, but before analyzing those
we use the algebraic gauge transformations in the last line of~\eqref{GAUGES} to set
\be\label{algeb}
\phi_\frak{Tr}=0\, ,\qquad \phi_{\wt{\frak{Div}}}=\frak{G} \chi.
\ee
After reaching this gauge choice there remain only residual gauge transformations with parameter $\varepsilon$ subject to
\be
\frak{Tr}\, \varepsilon = 0 = \wt{\frak{Div}}\, \varepsilon \, .
\ee
Note that in~\eqref{algeb}, having algebraically gauged away~$\phi_{\frak{Tr}}$, the pure trace part of~$\phi_{\wt{\frak{Div}}}$
cannot be removed via residual trace free gauge parameters $\varepsilon$; the remaining pure trace is parameterized by  the field~$\chi$. Having exhausted the algebraic gauge fixings, we now turn to the equations of motion.

The first equations of motion we look at are the algebraically solvable ones
corresponding to the two non-trivial commutation relations listed in Figure~\eqn{tractor-const}
\bea
\wt{\frak{Div}}\ \phi_{\frak{Grad}} - \frak{Grad}\ \phi_{\wt{\frak{Div}}}\ &=&\ \phi_{I\cdot D}\, ,\label{t1}\\[2mm]
\frak{Tr}\ \phi_\frak{Grad}-\frak{Grad}\ \phi_\frak{Tr}&=&2\, \phi_{{\frak{Div}}} \, .\label{t2}
\eea
Using the gauge choices~\eqref{algeb}, and in anticipation of its {\it r\^ole} as
the final physical field, calling
\be
\Phi\equiv \phi_\frak{Grad}\, ,
\ee
we have
\be
\phi_{I\cdot D}={\wt{\frak{Div}}}\ \Phi - \frak{G}\  \frak{Grad}\ \chi\, ,\qquad
\phi_{{\frak{Div}}}=\frac 12\, \frak{Tr}\ \Phi\, .
\ee
Now we must examine the remaining eight equations of motion;
some of these  are dynamical
but others are either dependent or amount to constraints on 
the physical fields. To begin with we examine
\be
{\frak{Div}}\  \phi_\frak{Tr} - \frak{Tr}\ \phi_{{\frak{Div}}}=0\, ,\label{t3}
\ee
which, in our gauge, gives the double trace condition 
\be
\frak{Tr}^2  \Phi=0\, ;
\ee
this is typical for higher spin theories.

In a similar vein, the equation of motion
\be
\wt{\frak{Div}}\  \phi_\frak{Tr} - \frak{Tr}\ \phi_{\wt{\frak{Div}}}=0\, ,\label{t4}
\ee
says
\be
\frak{Tr}\, \frak{G} \ \chi=0\, .
\ee
It is not difficult to verify that the operator~$\frak{Tr}\, \frak{G}$ is invertible so by algebraically solving this equation
of motion we learn
\be\ 
\chi=0\, .
\ee
Hence, the double tractor-trace free field $\Phi=\phi_{\frak{Grad}}$ is the only remaining physical field. It is subject to 
further constraints however, in particular the equations of motion
\bea
{\frak{Div}}\ \phi_{\wt{\frak{Div}}} - \wt{\frak{Div}}\ \phi_{{\frak{Div}}}&=&0\, ,\label{t5}\\[2mm]
\IcdotD \, \phi_{\wt{\frak{Div}}}-\wt{\frak{Div}}\ \phi_{I\cdot D}&=&0\, ,\label{t6}
\eea
give two further algebraic constraints
\be
\wt{ \frak{Div}}\, \frak{Tr} \ \Phi=\wt{\frak{Div}}{\ \!}^2\,  \Phi=0\, .
\ee
The final constraint comes from
\be
\frak{Grad}\ \phi_{{\frak{Div}}} - {\frak{Div}}\ \phi_\frak{Grad}=0\, ,\label{t7}
\ee
which says 
\be
\Big({\frak{Div}}-\frac12\, \frak{Grad}\, \frak{Tr}\Big)\Phi=0\, .
\ee
Note that although this constraint appears to be a differential one, at generic weights it can actually be solved algebraically
for lower slots of tractor fields because the top slot of the Thomas $D$-operator, $w(d+2w-2)$, does
not involve derivatives.

All other equations of motion are not independent save for 
\be
\IcdotD\, \phi_\frak{Grad}-\frak{Grad}\ \phi_{I\cdot D}=0\, ,\label{t8}
\ee
which gives the dynamical equation of motion
\be
\Big(\IcdotD- \frak{Grad}\, \frak{Div}\Big)\Phi=0\, .
\ee
The final set of tractor field equations are summarized in Figure~\ref{tequations}.
They agree with those of~\cite{Gover:2008pt,Gover:2008sw} save for the two constraints involving $\wt{\frak{Div}}$
which could not be detected by the spin two example explicitly checked there. The results above and  those of
Section~\bref{radial} therefore provide a proof of the conjecture of~\cite{Gover:2008pt,Gover:2008sw}.

In the next Section we employ ambient space BRST techniques to show that these equations indeed describe massive higher spin fields.

\subsection{BRST Derivation of On-Shell Massive Higher Spins}
\label{sec:BRST-ord}

We now consider the BRST operator $\Omega^{\rm ambient}$ implementing the complete set of ambient space constraints~\eqref{aconst}
\begin{equation}
\label{all-const}
\Omega^{\rm ambient}=\{\Delta,\,\, h,\,\,
\div, \,\,\tr,\,\,\IcdotD,\,\,\tilde\div, \,\,\grad\}
\end{equation} 
where all the constraints but $\grad$ are genuine constraints while $\grad$
implements a gauge symmetry. To simplify the exposition at this stage we do not
introduce new notations for the ghost variables but rely on the unambiguous assignments of the Grassmann parity,
ghost degrees and choice of representations given  in Section~\bref{sec:BRST-general}.

As we have seen in Section~\bref{sec:BRST-general}, if the constraints $\Delta,h$ are directly
imposed on the representation space, the BRST operator for the remaining
constraints  describes the gauge invariant equations of motion in
terms of tractors. 
Now we analyze another---ultimately equivalent---possibility. Namely, we show that by first directly  imposing
the constraints from another subalgebra, one ends up with the equations of motion for
massive higher spin fields on AdS space. As in Section~\bref{sec:amb-appr}
let us first use the constraints
$\IcdotD=I\cdot\dl{Y}=\frac{\partial}{\partial Y^{d+1}}$ and $\tilde\div=I\cdot
\dl{Z}=\frac{\partial}{\partial Z^{d+1}}$ (the third line of~\eqref{aconst}) to
eliminate the components $(Y^{d+1},Z^{d+1})$ of the coordinates~$(Y^M,Z^M)$ along the
scale tractor~$I^M$. This can be also done in BRST terms by reducing to the
cohomology of the part of the BRST operator containing these constraints.
Because this part is essentially a de Rham type differential it has no
cohomology classes depending on the respective {\rm ghost} variables and hence
the reduction just amounts to eliminating these components.

Hence, after taking into account the constraints $\IcdotD$ and $\tilde\div$  (and comparing~\eqn{d+1ops} and~\eqn{aconst})
we see that the theory is determined by the following BRST operator
\begin{equation}
\label{brst+tr}
 \Omega=c_0{\bf \Delta}+{\bf grad}\,  \dl{b}+c \, {\bf div}+\xi\, ({\bf tr}-2\dl{b}\dl{c})-c\dl{b}\dl{c_0}\,.
\end{equation} 
Here the ghost variables are all Grassmann odd and have ghost number assignments $\gh{c}=\gh{c_0}=\gh{\xi}=1=-\gh{b}$.
As the representation space we take the subspace of
functions of~$Y$ (defined on $(d+1)$-dimensional Minkowski space with the origin
excluded) taking values in polynomials of~$Z$ with the {\rm ghost} variables subject to
\begin{equation}\label{r-dep}
 \big(Y\, .\,  \dl{Y}-w+\ngh\big)\, \Psi=0\,,\mbox{ where } \ngh=c\dl{c}-b\dl{b}+2c_0\dl{c_0}\,.
\end{equation} 
This is just a BRST invariant extension of the $(Y.\dl{Y}-w)\Psi=0$ constraint introduced before. 

Our claim is that the equations of motion~$\Omega\, \Psi=0$, subject to the gauge
invariance~$\Psi\sim\Psi+\Omega \, \Xi$, where~$\gh{\Psi}=0$ and  $\gh{\Xi}={-1}$,
describe massive, constant curvature,  higher spin fields. More precisely, to describe
a field of definite spin~$s$ one  imposes the constraint:
\begin{equation}
 N_s\Psi=0\,,\qquad N_s:=Z\, .\, \dl{Z}+b\dl{b}+c\dl{c}+2\xi\dl{\xi}-s\,.
\end{equation} 

This is consistent because $N_s$ commutes with both the BRST
operator~\eqref{brst+tr} and the constraint~\eqref{r-dep2}. Note that, as we
have alredy discussed in Section~\bref{log R} in general, this still may not
describe an irreducible theory because of the genuine gauge invariance present
in the theory for special values of $w$ that correspond to (partially) massless
fields. To describe an irreducible system in those cases one needs to impose extra
irreducibility conditions $(Y.\dl{Z})^t\, \Psi=0$ where $t$ is the depth\footnote{This is a straightforward
generalization~\cite{Alkalaev:2009vm} to the partially massless case of the massless irreducibility condition
$Y.\dl{Z}\Psi=0$ from~\cite{Fronsdal:1978vb}.}, see
Section~\bref{sec:parent-ads} as well as the \hyperlink{decouples}{discussion in Section~\bref{log R}}.

As the first step towards a proof of our claim, we further simplify the formulation by imposing the
{\rm ghost}-extended trace constraint directly on states rather then keeping it
in the BRST operator. This is legitimate because the nilpotent term~$\xi\, ({\bf
tr} -2 \dl{c}\dl{b})$ is algebraic and has no~$\xi$-dependent cohomology
classes. After this the system is defined by the following BRST operator
\begin{equation}
\label{atheor}
%\begin{gathered}
\widehat\Omega=c_0{\bf \Delta}+{\bf grad} \, \dl{b}+c \, {\bf div} -c\dl{b}\dl{c_0}\,,
\end{equation}
while $\Psi$ is assumed to satisfy
\begin{equation}
\label{atheor-const}
 \big(Y\, .\, \dl{Y}-w+\ngh\big)\, \Psi=0\, , \qquad  \big({\bf tr}-2\dl{c}\dl{b}\big)\, \Psi=0\,.
%\end{gathered}
\end{equation}
This BRST operator along with the trace constraint is
known~\cite{Bengtsson:1986ys,Ouvry:1986dv} (see also~\cite{Henneaux:1987cp}) to describe massless higher spin gauge fields on~$(d+1)$-dimensional Minkowski space. The remaining constraint
consistently eliminates the radial dependence of fields hence giving massive,
constant curvature, fields.

A detailed proof of the above statement uses a more advanced technique that will be
given  in Section~\bref{sec:parent-ads}, but the underlying
mechanism is  simple so we explicitly give the main arguments here. One first takes
a specially adapted coordinate system: dilation-invariant coordinates~$x^\mu$
and a radial coordinate~$r=\exp(u)$ (see the metric~\eqn{radialds2}). Then the
analysis is  rather similar to that presented in Section~\bref{log R}, except
that our aim is to directly obtain the onshell conditions~\eqn{onshell}. Let us
note however that the formulation considered in Section \bref{log R} can be also
derived immediately from~\eqref{atheor},~\eqref{atheor-const} by reducing to the
cohomology of cubic ghost term $c\dl{b}\dl{c_0}$ (the reduction is algebraically
identical to its flat space version considered in~\cite{\BGST}).

Because of the constraint~\eqref{r-dep} and equation~\eqn{euler},
the~$u$-dependence is completely fixed so that the field theory is now
 defined on the hyperboloid~$u=0$. The gauge transformation
determined by ${\bf grad}$, as given in~\eqn{grad2grad}, contains the
term~$Z^{(u)} \dl{u}$, but thanks to~\eqn{r-dep}, on the
hyperboloid~$\dl{u}\Psi=(w-\ngh)\Psi$. This implies that for a generic value
of~$w$ one can completely eliminate any dependence on~$Z^{(u)}$. In other words,
in this case the gauge symmetry is an auxiliary St\"uckelberg one and suffices 
to eliminate the log-radial oscillator~$Z^{(u)}$  (or in terms of the $Y^A,Z^A$
coordinates, the gauge condition $Y^A\dl{Z^A}\Psi=0$ is reachable and
completely removes the gauge freedom). For the special case of spin~2, one can
see how the above argument works from the explicit formula for the gauge
transformations~\eqn{stg}. 

As a result of this elimination
both the variables~$Z^{(u)}$ and the {\rm ghost}~$b$ are removed. There are no more variables of negative {\rm ghost} degree so that
the resulting theory is non-gauge and the equations of motion are just the remaining constraints. In the absence
of the~$Z^{(u)}$ and~$b$-variables these are given by
\begin{equation}\label{consts}
 {\bf \Delta}\, \varphi=0\,, \qquad {\bf div}\, \varphi=0\,, \qquad {\bf tr}\, \varphi=0\, , 
\end{equation} 
where $\varphi=\varphi(u,x^\mu;Z^m)$. 

It is easy, following the explanation of the log radial reduction given in Section~\bref{log R},
to write out equations~\eqn{consts} in terms of $d$-dimensional AdS operators, in particular
\be\nn
{\bf tr}\ ={\rm tr}-\big[\dl{Z^{(u)}}\big]^2\, ,
\ee
\be\nn
{\bf div}=e^{-u}\big({\rm div} - \big[\frac{\partial}{\partial u} +{\rm N}+d\big]\frac{\partial}{\partial Z^{(u)}}-Z^{(u)}\,  {\rm tr}\big) \, ,
\ee
which immediately implies ${\rm tr}\, \varphi={\rm div}\, \varphi=0$. Finally, from ${\bf \Delta}=[{\bf div}, {\bf grad}]$, 
along with the expression for ${\bf grad}$ in~\eqn{grad2grad} as well as the algebra~\eqn{algebra}  we learn that
\be
\label{mass-w}
(\nabla^2+s)\,  \varphi = w(w+d-1)\, \varphi \, .
\ee
Comparing with~\eqn{onshell} we see that the system~\eqn{consts} describes a massive, spin~$s$ AdS field with mass
given by~\eqref{mass}.

This relation for the mass also follows from a representation theoretic argument.
Namely representing~$\nabla^2$ through the orbital part~$L_{AB}=Y_A\dl{Y^B}-Y_B\dl{Y^A}$ of the AdS generators
via~\cite{Pilch:1984xx}
(see also~\cite{Metsaev:1995re,deWit:2002vz,Deser:2003gw})
\begin{equation}
\nabla^2+s=-\half L^{AB}L_{AB}=-Y^2\,{\bf \Delta}+Y.\dl{Y}\Big(Y.\dl{Y}+d-1\Big)\, ,\,
\end{equation} 
and identifying the radial conformal isometry with the weight $w$, we again obtain~\eqn{mass-w}.

Our final computation, is a parent formulation of the BRST approach presented above. This is a useful method for
analyzing the gauge invariances of the system including its
partially massless limits. It also provides a method to construct an unfolded formulation.

\subsection{Parent BRST Formulation of Tractor Fields}
\label{sec:parent-tractor}

Following~\cite{Barnich:2006pc} (see also \cite{Bekaert:2009fg,Alkalaev:2009vm} for a more recent discussion)
we now represent the theory defined by~\eqref{all-const} in a first order form with respect to the  intrinsic
geometry of the AdS space while keeping  covariance under  $\mathfrak{o}(d,2)$. 
This method amounts to treating the representation
space as a fiber over the true AdS space~$\manM$ and gluing  fibers together
with the help of an appropriate covariant derivative that enters the formalism
as an extra term in the BRST operator.

More technically, we consider the vector bundle over $\manM$ associated with the
flat version of the tractor bundle described in Section~\bref{gen-trac} and with fiber being
the ambient space itself extended by the oscillator variables~$Z^M$. The resulting theory is
then determined by the following BRST operator\begin{equation}
\label{ambient-parent}
\Omega^{\rm amb.~parent}= \nabla+\overline\Omega^{\rm \,ambient},
\end{equation} 
where
\begin{equation}
\label{covariant}
\nabla:=\theta^\mu\dl{x^\mu}-
\theta^\mu {\mathcal A} _\mu{}^N{}_{M}\Big((Y^M+X^M)\dl{Y^N}+Z^M\dl{Z^N}\Big) 
\end{equation}
is the flat covariant derivative originating from that in the tractor bundle, and
$\overline\Omega^{\rm \,ambient}$ is the BRST operator corresponding to the
constraints~\eqref{all-const} which are  now implemented in the fiber.
Here~$x^{\mu}$ are local coordinates on~$\manM$ and a suitable local frame such
that $X^M=\const$ is chosen. In addition, basis differential forms~$dx^\mu$ are
replaced with extra Grassmann odd {\rm ghost} variables~$\theta^\mu,$~$\;\mu=0,
... , d-1$, because~$\nabla$ now appears as a part of the BRST operator.

The BRST operator~\eqref{ambient-parent} also involves the scale tractor $I^M$
through the constraints~\eqref{aconst}. This does not spoil nilpotency because
$\nabla I=0$ by assumption. The representation space is chosen to be functions
of $(x,\theta)$ tensored with the representation space for the fiber part of the BRST operator, 
$\bar\Omega^{\rm ambient}$. The latter is identical to that of the previous
Section except that we now represent the variables $Y^M$ in the expression for
$\bar\Omega^{\rm ambient}$ differently. Namely, $Y$ and $\dl{Y}$ act on the
space of formal power series in $Y$ according to $Y^M \to Y^M+X^M$ and
$\dl{Y^M}\to \dl{Y^M}$.  For instance, the constraint $h$ from~\eqref{aconst}
fixing the radial dependence acts as $(Y^M+X^M)\dl{Y^M}-w$.
Note that this ``twisted'' representation (introduced originally in~\cite{\BGadS}) is
inequivalent to the usual one because the redefinition $Y^M+X^M \to Y^M$ is not
defined for generic formal power series in $Y^M$.

More generally, this formulation can be seen as a Fedosov-type~\cite{Fedosov:1994} extension of the
BRST constrained system of the previous Section; it can also be related to an unfolded
formulation for conformal fields~\cite{Vasiliev:2009ck}. 
We refrain from giving an extensive discussion and refer instead
to~\cite{\BGST,Barnich:2006pc,Bekaert:2009fg,Alkalaev:2009vm}. Note, however, that in
contrast to the analogous BRST operator considered in~\cite{Bekaert:2009fg}, the operator 
\eqref{ambient-parent} is not conformally invariant because it explicitly involves the
scale tractor $I$ which breaks manifest ${\mathfrak o}(d,2)$-invariance down to
${\mathfrak o}(d-1,2)$. Also note that, specializing the arguments given
in~\cite{\BGST,\BGadS,Bekaert:2009fg,\AG} to the case at hand, it follows that
the theory determined by $\Omega^{\rm parent}$ is equivalent to that determined
by the BRST operator~$\Omega^{\rm ambient}$. 

The model determined by the BRST operatror~\eqref{ambient-parent} serves as a parent theory
for all the formulations considered above. In particular, to obtain the formulation in terms of tractors
one should reduce to the cohomology of the part of the BRST operator
containing ghosts $\theta^\mu$ and constraints ($\Delta$,
$(Y+X)\cdot\dl{Y}-w$). This allows us to eliminate all the $Y^A$-variables so the resulting 
$Y$-independent fields can be identified with the symmetric
tractors from Section~\bref{OSC}.

More precisely, for $w$ generic, the space of sections $\Phi(x,Z,Y)$ satisfying 
\begin{equation}
\label{3const}
\nabla\Phi=0 \,, \qquad \big((Y+X)\cdot\dl{Y}-w\big)\Phi=0\,, \qquad \Delta\Phi=0\, , 
\end{equation} 
can be shown to be isomorphic to the space of symmetric weight $w$ tractors. Indeed, in this case
any $\phi(x,Z)$ can be uniquely extended to $\Phi(x,Z,Y)$ satisfying~\eqref{3const}
and $\Phi|_{Y=0}=\phi$. 

Now, expanding
$$
\Phi= \phi + \phi_M Y^M + \phi_{MN} Y^M Y^N+\cdots\, ,
$$
at leading and next to leading order $\nabla\Phi=0$ says
$$
\phi_\mu=\nabla_\mu\phi\, ,\qquad
\phi_\mu{}^M=\frac{1}{2}\nabla_\mu \phi^M=\frac12 \begin{pmatrix}
\nabla_\mu \phi^+-\phi_\mu\\[1mm] \nabla_\mu \phi^m +e_\mu{}^m \phi^- +\Rho_\mu^m \phi^+\\[1mm]
\nabla_\mu \phi^--\Rho_\mu^m \phi_m
\end{pmatrix}\, ,
$$ 
where $\nabla_\mu=\d_\mu-{\cal A}_{\mu}{}^M{}_N Z^N\dl{Z^M}$ on the (far) right hand side of each equality is the tractor covariant derivative
acting on totally symmetric tractors. Similarly, from the second relation in~\eqn{3const} we learn
$$
\phi^+=w\phi\, ,\quad \phi^{M+}=\frac12(w-1)\phi^M\, .
$$
Finally, the harmonic condition in~\eqn{3const} yields
$$
\phi^-=-\ \frac{g^{\mu\nu}\nabla_\mu\nabla_\nu+w\, \Rho}{d+2w-2}\, \phi\, ,\qquad
$$
Orchestrating the above computations gives
\begin{equation}
\label{phi-Phi}
\Phi(x,Z,Y)=\phi(x,Z)\ +\ \frac{Y^M D_M\phi(x,Z)}{d+2w-2}\ +\ O(Y^2)\, ,
\end{equation}
where $D_M$ is precisely the Thomas $D$-operator (acting on totally symmetric weight~$w$ tractor tensors) defined in~\eqn{ThomD},
so we are really working with weight $w$ tractors.

For the above identification to work for all values
of $w$ it is better to replace the  condition $\Delta \Phi=0$ in~\eqref{3const} by the equivalence relation
({\it cf.} equation~\eqref{equiv})
\begin{equation}
\Phi(x,Z,Y) \sim \Phi(x,Z,Y)+(Y+X)^2\, \chi(x,Z,Y)\,.
\end{equation} 
Note that a closely related representation was proposed in~\cite{Bekaert:2009fg} to 
describe certain conformal fields.

To demonstrate explicitly how the ambient gauge transformation \begin{equation}\delta
\psi=Z\cdot \frac{\partial\chi}{\partial Y}\, ,\end{equation}  
produces tractor gauge transformations, we consider~$\Phi$ in~\eqref{phi-Phi} depending on the 
ghost variable $b$ and adjust the second
constraint in~\eqref{3const} accordingly as
\begin{equation}\big((Y+X)\cdot\dl{Y}-b\dl{b}-w\big)\Phi=0\, .\end{equation} 
With this adjustment the term
$Q:=Z\cdot\dl{Y}\dl{b}$ from the BRST
operator~\eqref{ambient-parent} acts in the subspace~\eqref{3const}. When the gauge parameter
$\Xi(x,Z,b,Y)=b\, \xi(x,Z)+\cO(Y)$ and the field $\Phi(x,Z,Y)=\phi(x,Z)+\cO(Y)$
both satisfy the constraints~\eqref{3const}, then~\eqref{phi-Phi} implies that the gauge
transformation $\delta\Phi=Q\, \Xi$ in terms of $\phi$ and $\xi$ takes the form
$\delta_\xi\phi=Z^M D_M \xi=\mathfrak{Grad} \, \xi$ in concordance with Figure~\ref{tequations}.

In fact one can show that the parent formulation
determined by~\eqref{ambient-parent} can be consistently reduced to the
formulation determined by $\Omega^{\rm inter}=\nabla+Q$ where all the constraints save for
$\mathrm{Grad}$ are imposed on states and all the associated ghost variables are
set to zero. This formulation is constructed and utilized in the next Section (see Eq.~\eqref{inter}).

In general, the treatment of tractor bundles through~\eqref{3const} beyond 
conformally flat spaces requires that one extends $\nabla$ in~\eqref{3const} by terms
nonlinear in $Y$  in order to maintain nilpotency, $\nabla^2=0$.
Such an extension can in fact be considered as a conformal geometry version of
the Fedosov~\cite{Fedosov:1996fu} connection. Note that similar extensions of the 
AdS and flat space connections were considered
in~\cite{Grigoriev:2006tt,Vasiliev:2005zu}. The identification of the Fedosov connection
as a BRST operator was discussed in~\cite{Grigoriev:2000rn} (see
also~\cite{Batalin:2001je,\BGST,Grigoriev:2006tt}).
Another related way to view the parent formulation of tractor fields is to
interpret it as a conformally covariant extension of a jet bundle.
In this context it is worth mentioning the rather elegant conformally covariant tensor calculus of~\cite{Boulanger:2004eh} based on a jet bundle approach. In fact, the relationship between the tensor
calculus of~\cite{Boulanger:2004eh} and the parent BRST formulation can be
established using the general framework of~\cite{Barnich:2010sw}.

The equations~\eqref{3const} also provide an algebraic framework to analyze
special values of the weight $w$. Indeed, solving~\eqref{3const} with the
initial condition $\Phi(x,Y,Z)|_{Y=0}=\phi(x,Z)$ is in fact equivalent to
finding an ambient extension of $\phi(x,Z)$ defined on an appropriately gauge-fixed
submanifold of the null-cone in the ambient space. This equivalence can be explicitly
seen by applying the arguments from~\cite{Alkalaev:2009vm} (see also
\cite{\BGadS,Bekaert:2009fg}). In fact,  it is not difficult to prove that a solution for the extension exists for arbitrary $\phi$
unless $d+2w=2l$ with $l$ a positive integer as was already discussed in
Section~\bref{ambient space}.

\subsection{Parent BRST Formulation in Terms of AdS fields}
\label{sec:parent-ads}

Starting from the parent formulation of the previous Section one can also arrive at a
description of massive AdS fields similar to the one developed
in~\cite{\BGadS,\AG} in the case of (partially) massless fields. 
The first steps of
this reduction are nearly identical to those of the reduction considered
in Section~\bref{sec:BRST-ord}. We chose a scale and a local frame  such 
that 
$I^{d+1}=1$, $I^{M\neq d+1}=0$, $X^d=X^{d+1}=1$, $X^{M\neq d,d+1}=0$, with
 metric  given by~\eqn{flatamb}. As in Section~\bref{sec:amb-appr} and \bref{sec:BRST-ord},
the  constraints $I\cdot\dl{Y}$ and $I\cdot\dl{Z}$ can be used to eliminate the 
fiber  coordinates $Y^{d+1}$ and $Z^{d+1}$ in the direction of the scale
tractor~$I$. This can be also seen as a reduction to the cohomology of the part
of the BRST operator containing these constraints.

The next step is to  impose the ghost-extended constraint  $(Y+X)\cdot\dl{Y}-w$ directly on the representation space rather than
taking it into account in the BRST operator. Again, this can be seen as a
reduction to the cohomology of the term in the BRST operator containing this
constraint. The resulting formulation is then described by a BRST operator with the 
same structure  as~\eqn{ambient-parent}
\begin{equation}
\label{embed-parent}
\Omega^{\rm parent}= \nabla+\overline\Omega,
\end{equation} 
where~$\overline\Omega$ is now the BRST operator~\eqref{brst+tr} acting on the fiber, and
\begin{equation}
\label{embed-parent-nabla}
\nabla=\theta^\mu\dl{x^\mu}-
\theta^\mu \omega_\mu{}^A{}_{ B}\Big((Y^B+V^B)\dl{Y^A}+Z^B\dl{Z^A}\Big) 
\end{equation}
is the restriction of the covariant derivative~\eqref{covariant} to  $Y^{d+1}$ and $Z^{d+1}$-independent sections.

The geometry underlying this reduction of the covariant derivative
is as follows: Restricting to $Y^{d+1}$ and $Z^{d+1}$-independent elements corresponds to 
a quotient of the $(d+2)$-dimensional fiber by the one-dimensional subspace
generated by the direction of the scale tractor~$I$.  Indeed, $Y^{d+1},Z^{d+1}$-independent elements are exactly
those constant along the subspace
generated by $I$ and can therefore  be seen as functions on the quotient. The ${\mathfrak o}(d,2)$ connection reduces then to an ${\mathfrak o}(d-1,2)$
one, where ${\mathfrak o}(d-1,2)\subset {\mathfrak o}(d,2)$ is the subalgebra stabilizing $I$.
The components of the reduced connection are just the original
components $\mathcal{A}_\mu{}^M{}_{ N}$, where the indices $M,N$ are restricted to take values from $0$ to
$d$.

Under this reduction, the term in the covariant derivative involving the section~$X$
produces a term of the same structure but with $X$ replaced by a section $V$ of
the reduced bundle. At any point the section $V$ is simply a projection of $X$ to
the quotient space so that in components one has $V^M=X^M$ for $M\neq d+1$. In
particular, $X^2=0$ implies $V^2=-1$. Because $I$ is covariantly
constant, the rank of $e_\mu{}^A=\nabla_\mu V^A$ is also maximal (recall that
we required $\nabla X$ to have maximal rank) so one can identify the
reduced bundle with the vector bundle used to describe AdS
fields~\cite{MacDowell:1977jt,Stelle:1979aj} (see
also~\cite{Vasiliev:2001wa,Bekaert:2005vh,\BGadS,\AG} for a more recent and
closely related treatment). In that context, the ${\mathfrak o}(d-1,2)$-connection and the
section $V$ satisfying $V^2=-1$ (known as a compensator) were used to describe
the AdS geometry.

The formulation based on~\eqref{embed-parent} and \eqref{embed-parent-nabla} has ${\mathfrak o}(d-1,2)$
invariance manifestly realized. Moreover, it is well suited for studying the structure of the gauge symmetries
of the theory. To that end, let us concentrate on the constraint fixing the radial dependence
\begin{equation}
\label{r-dep2}
 \left((Y+V). \dl{Y}-w+\ngh\right)\Psi=0\,.
\end{equation} 
Suppose that a local frame is chosen such that~$V=(0,\ldots,0,1)$ with fiberwise metric
$\eta_{AB}={\rm diag}(-1,1,\ldots,-1)$. We also decompose~$Y^A=(y^a,\overline y)$, $Z^A=(z^a,\overline z)$. 
It is now easy to show that the radial oscillator~$\overline z$ and the {\rm ghost} variable~$b$
can be consistently eliminated. Indeed, one first solves the constraint~\eqref{r-dep2}
as a formal power series in $\overline y$. The solution can be compactly written as
\begin{equation}
 \Psi(y,\overline y, Z,{\rm {\rm ghost}s})
=(\overline y+1)^{{}^{w-y^a\dl{y^a}-N_{{\rm ghost}}}}\psi(y,Z,{\rm {\rm ghost}s})\, ,
\end{equation} 
where the operator in front of $\psi$ is understood as a formal power series in $\overline y$
(note that $y^a\dl{y^a}+N_{{\rm ghost}}$ acts as multiplication by a number on any homogeneous term, see~\cite{\BGadS} for more details).
This establishes a one to one  correspondence between elements satisfying~\eqref{r-dep2} and $\overline y$-independent elements. 

Because the term ${\bf grad}\dl{b}$ in $\overline\Omega$ determining the gauge
transformation commutes with the operator on the left hand side of~\eqref{r-dep2}, its action can be represented in
terms of $\overline y$-independent elements where it acts as
$\hat{\bf grad}\dl{b}$ with
\begin{equation}
 \hat{\bf grad} =z^a\dl{y^a}+\overline z \, (w-y^a\dl{y^a}-N_{{\rm ghost}})\,,
\end{equation} 
so that the respective gauge symmetry is indeed algebraic for~$w$ generic.

It turns out that the dependence on $\bar z$ and $b$ can be eliminated thanks to the last term
in the above expression for $\hat{\bf grad}$:  Taking as  degree the difference between the homogeneities in~$\theta^\mu$
and $\overline z$ (because only polynomials in~$Z$ are allowed the degree is bounded from below) the lowest degree term in~$\Omega^{\rm parent}$ is
\begin{equation}
\label{st1}
 \Omega_{-1}=(w-y^a\dl{y^a}-N_{{\rm ghost}}+1) \, \overline z\,  \dl{b}\, ,
\end{equation} 
where we made use of $\commut{N_{\rm ghost}}{\dl{b}}=\dl{b}$.
If~$w$ is generic, the cohomology representatives can be chosen
to be~$\overline z, b$-independent. Because the cohomology of~$\Omega^{\rm parent}$ is isomorphic to
that of~$\hat\Omega$ from~\eqref{atheor}, this gives a rigorous and purely algebraic realization of the argument
used in Section~\bref{sec:BRST-ord} to show that the radial oscillator and the ghost $b$
can be eliminated for $w$ generic.

In this case the reduced BRST operator (obtained after eliminating the variables $\overline z$ and $b$) has the structure
\begin{equation}
\Omega_{\rm red}=\hat\nabla+c\,\hat{ \bf div}+c_0 \,\hat {\bf \Delta}\, ,
\end{equation} 
where hats over operators indicate that they are reduced to
$\Omega_{-1}$ cohomology in the subspace~\eqref{r-dep2}. As there are no
variables of negative ghost degree left, there are no gauge invariances and
the physical fields are represented by ghost-independent elements so that the
equations of motion determined by $\Omega_\red$ are just
\begin{equation}
 \hat\nabla\psi(x,y,z)=0\,, \qquad  \hat{\bf div}\, \psi=\hat{\bf \Delta}\, \psi=0\,,
\end{equation} 
(recall also the tracelessness condition).
These equations are an unfolded version of the massive
equations of motion. Indeed, the second and the third equations are 
algebraic constraints specifying the subspace (known as the Weyl module in the unfolded formalism)
where $\psi$ takes values  while the first one has the form of a
covariant constancy condition. We do not go into further details and refer
instead to~\cite{Ponomarev2010}, where the unfolded description of massive
totally symmetric fields on AdS has been constructed.

If $w$ is not generic the gauge properties of the model are, of course, more subtle.
Let us show how the parent formulation handles this situation:
We return to   the formulation~\eqref{embed-parent},~\eqref{embed-parent-nabla},
with the constraint $(Y+V)\cdot\dl{Y}-w+\ngh$ explicitly imposed,
and reduce it to the cohomology of the following term of the BRST operator~$\Omega^{\rm parent}$
\begin{equation}
c_0{\bf \Delta}+c\, {\bf div}+\xi\, {\bf tr}\,.
\end{equation}
Note that this and the following reductions are  exactly the same
as in~\cite{\BGadS} to which we refer for details.
The cohomology is known and can be identified with the subspace of totally traceless (in both $Z$ and $Y$ spaces) and $c_0,c,\xi$-independent elements.  

The reduced theory is then determined by the BRST operator 
\begin{equation}
\label{inter}
 \Omega^{\rm inter} =\nabla + Q\,, \qquad Q:={\bf grad}\dl{b}\,,
\end{equation} 
where states $\Psi(x,Y,Z,b,\theta)$ are assumed to be totally traceless and are subject to the constraint $((Y+V)\cdot\dl{Y}-w-b\dl{b})\Psi=0$.
As a next step one can reduce to the cohomology of the second term $Q$ in $\Omega^{\rm inter}$ (which results in an unfolded version of the system). In general there can be both $b$-dependent
and $b$-independent cohomology classes. When $b$-dependent classes are absent then there are no elements of negative ghost
degree and the theory is non-gauge (as we have already seen for generic $w$). 

Therefore we consider the cohomology of the second term at ghost degree $-1$ ({\it i.e.}, we are looking for $b$-dependent classes).
For $\Psi=b\,  \phi$ the cocycle condition and  constraints give
\begin{equation}
\label{3}
Z\, .\, \dl Y \phi=0\,, \quad  ((Y+V)\, .\, \dl{Y}-w-1)\phi=0\,, \quad (Z\, .\, \dl{Z}-s+1)\phi=0\, ,
\end{equation} 
where we have skipped the three trace constraints and explicitly added the constraint
singling out a spin~$s$ field. (Recall that to describe a spin-$s$ field in 
the formulation~\eqref{embed-parent}, \eqref{embed-parent-nabla}, one must in addition 
impose the  constraint $N_s=Z^A\dl{Z^A}+b\dl{b}+c\dl{c}+2\xi\dl{\xi}-s$. This gives the third equation in~\eqref{3}.)

The first constraint in~\eqref{3} tells us that the homogeneity degree in $Z$ is
greater or equal to that in $Y$ for any homogeneous component of $\phi$ so
that $\phi$ is polynomial in both $Y$ and $Z$ (indeed, this is simply a Young
condition for the respective Young tableaux). In terms of $Y^\prime=Y+V$ the
second condition takes the form $(Y^\prime \cdot\dl{Y^\prime}-w-1)\phi=0$ and obviously
has  solutions only for $w\geq -1$ and integer. Moreover, the above argument
also shows that $w+1\leq s-1$ so that $w \leq s-2$. One then concludes that
the residual gauge invariance is present only for $w=-1,0,\ldots,s-2$. For other
values of $w$ the cohomology at ghost degree $-1$ is empty and the theory
is non-gauge. In particular, for generic $w$, this reproduces the
analysis in the beginning of this Section.

Let us concentrate now on the gauge invariant case. Suppose that $w$
is integer and satisfies $-1 \leq w\leq s-2$. In this case there
are nontrivial cohomology classes at ghost degree $-1$ and hence
genuine gauge fields. For instance if ~$(Z\dl{Z}+b\dl{b}-s)\Psi=0$, {\it i.e.},
if we are describing a spin~$s$ field, let us take~$w=s-t-1$ with $t=1,\ldots,s$. In this case the
system describes a partially massless field of depth~$t$ along with some extra (though decoupled) degrees of freedom. By
imposing in addition a constraint~$[(Y+V)\dl{Z}]^t$ one singles out just the
irreducible field. This formulation of partially massless fields was developed
in~\cite{Alkalaev:2009vm}. By explicitly reducing the formulation to the cohomology of ${\bf grad}\dl{b}$
one can arrive at the unfolded form of the partially massless fields originally 
proposed in~\cite{Skvortsov:2006at}.

As a final remark let us return to the analysis of the cohomology at ghost number
zero for special values of $w$. Consider the cohomology of the operator~\eqref{st1}. 
As we have already seen for $w$ non-integer or $w<-1$, its
cohomology is given by $\bar z,b$-independent elements. It is easy to see that
$w\geq -1$ and integer are special values because the $\bar z$-dependence can not be
completely eliminated. As we have seen, weights $-1 \geq w \geq s-2$ and integer
correspond to (partially) massless fields and in this case there is a genuine
gauge freedom and an extra irreducibility condition is needed. For $w>s-2$ and
integer there are no gauge fields but still the $\bar z$-dependence can not be completely
eliminated.
This signals that for such values of $w$ the structure of the space of solutions
to the equations of motion can be different. However this requires further study
which we leave for  future work.

\section*{Acknowledgments}
 A.W. is indebted to the Lebedev Physics Institute for hospitality.
M.G. is grateful to G.~Barnich, X.~Bekaert, E.~Skvortsov and especially to K.~Alkalaev and R.~Metsaev for useful discussions. The work of M.G. was supported by RFBR grant 11-01-00830.

\addtolength{\baselineskip}{-3pt}
\addtolength{\parskip}{-3pt}

\providecommand{\href}[2]{#2}\begingroup\raggedright\endgroup

\end{document}